\documentclass{aa}

\usepackage[varg]{txfonts}
\usepackage{natbib}
\usepackage{blindtext}
\usepackage{booktabs}
\usepackage{pifont}
\usepackage{lscape}
\usepackage{multirow}
\usepackage{tabularx}
\usepackage{pbox}
\usepackage{wasysym}
\usepackage{cleveref}
\usepackage{upgreek}
\usepackage{listings}
\usepackage{color}

\newcommand{\nha}{\ensuremath{N_\text{H,1}}\xspace}
\newcommand{\nhb}{\ensuremath{N_\text{H,2}}\xspace}
\newcommand{\nh}{\ensuremath{N_\text{H}}\xspace}
\newcommand{\fcov}{\ensuremath{f_\text{cvr}}\xspace}

\newcommand{\swift}{\textit{Swift}\xspace}
\newcommand{\xmm}{\textit{XMM-Newton}\xspace}

\newcommand{\suzaku}{\textit{Suzaku}\xspace}
\newcommand{\integral}{\textit{INTEGRAL}\xspace}

\newcommand{\chandra}{\textit{Chandra}\xspace}
\newcommand{\hst}{\textit{HST}\xspace}
\newcommand{\nustar}{\textit{NuSTAR}\xspace}

\begin{document}

\title{A \textit{Suzaku}, \textit{NuSTAR,} and \textit{XMM-Newton} view
  on variable absorption and relativistic reflection in NGC~4151}

\author{T.~Beuchert\inst{\ref{inst1},\ref{inst2}}
\and A.G.~Markowitz\inst{\ref{inst3}}
\and T.~Dauser\inst{\ref{inst1}}
\and J.A.~Garc\'ia\inst{\ref{inst4},\ref{inst5},\ref{inst1},\ref{inst6}}
\and M.L.~Keck\inst{\ref{inst7}}
\and J.~Wilms\inst{\ref{inst1}}
\and M.~Kadler\inst{\ref{inst2}}
\and L.W.~Brenneman\inst{\ref{inst5}}
\and A.A.~Zdziarski\inst{\ref{inst8}}
}

\institute{Dr.\ Remeis-Observatory \& Erlangen Centre for Astroparticle
  Physics, Universit\"at Erlangen-N\"urnberg, Sternwartstrasse 7,
  96049 Bamberg, Germany 
\label{inst1}
\and Lehrstuhl f\"ur Astronomie, Universit\"at W\"urzburg, 
Emil-Fischer-Straße 31, 97074, W\"urzburg, Germany
\label{inst2}
\and Center for Astrophysics and Space Sciences, University of
California, San Diego,  9500 Gilman Dr., La Jolla, CA 92093-0424, USA
\label{inst3}
\and Cahill Center for Astronomy and Astrophysics, California
Institute of Technology, Pasadena, CA 91125, USA
\label{inst4}
\and Harvard-Smithsonian Center for Astrophysics, 
60 Garden St., Cambridge, MA 02138, USA
\label{inst5}
\and Alexander von Humboldt Fellow
\label{inst6}
\and Institute for Astrophysical Research, Boston University,
725 Commonwealth Avenue, Boston, MA 02215, USA
\label{inst7}
\and Nicolaus Copernicus Astronomical Center, Polish Academy of
Sciences, Bartycka 18, PL-00-716 Warsaw, Poland
\label{inst8}
}

\date{xx-xx-xx /
xx-xx-xx }

\abstract {We disentangle X-ray disk reflection from complex
  line-of-sight absorption in the nearby Seyfert NGC~4151, using a
  suite of \textit{Suzaku}, \textit{NuSTAR}, and \textit{XMM-Newton}
  observations.  Extending upon earlier published work, we pursue a
  physically motivated model using the latest angle-resolved version
  of the lamp-post geometry reflection model \textsc{relxillCp\_lp}
  together with a Comptonization continuum.  We use the long-look
  simultaneous \textit{Suzaku}/\textit{NuSTAR} observation to develop
  a baseline model wherein we model reflected emission as a
  combination of lamp-post components at the heights of 1.2 and 15.0
  gravitational radii. We argue for a vertically extended corona as
  opposed to two compact and distinct primary sources.  We find two
  neutral absorbers (one full-covering and one partial-covering), an
  ionized absorber ($\log \xi = 2.8$), and a highly-ionized ultra-fast
  outflow, which have all been reported previously. All analyzed
  spectra are well described by this baseline model.  The bulk of the
  spectral variability between $\sim$1\,keV and $\sim$6\,keV can be
  accounted for by changes in the column density of both neutral
  absorbers, which appear to be degenerate and inversely correlated
  with the variable hard continuum component flux.  We track
  variability in absorption on both short (2\,d) and long
  ($\sim$1\,yr) timescales; the observed evolution is either
  consistent with changes in the absorber structure (clumpy absorber
  at distances ranging from the broad line region (BLR) to
  the inner torus or a dusty radiatively driven wind) or a
  geometrically stable neutral absorber that becomes increasingly
  ionized at a rising flux level. The soft X-rays below 1\,keV are
  dominated by photoionized emission from extended gas that may act as
  a warm mirror for the nuclear radiation.}


\maketitle

\section{Introduction}\label{sec:intro}
Active galactic nuclei (AGN) efficiently return energy to their
environment both via accretion onto supermassive black holes (SMBHs)
resulting in broad-band radiation and via the ejection of matter
through collimated jets and outflows. Jets can form by extracting
energy from the inner edge of an accretion disk around a rotating Kerr
black hole \citep{BZ1977,Tchekhovskoy2011} while a maximally spinning
black hole allows the innermost stable circular orbit (ISCO) of the
disk to lie very close to the black hole. A hot electron plasma of yet
unclear geometry, the so-called ``corona'', is supposed to upscatter
soft thermal seed photons from the accretion disk
\citep{Haardt1993,Dove1997a,Dove1997b,Belmont2008}
approximately resulting in an X-ray power-law spectrum that can be observed both
directly and reflected off the disk. For radio-quiet AGN in the absence of
strong jets, the X-ray spectrum is entirely dominated by emission
processes from the compact X-ray-emitting regions. The
prominent fluorescent Fe~K$\alpha$ line at a rest-frame
energy of 6.4\,keV as well as a broad Compton hump peaking around
20--30\,keV are well explained by distant reflection
\citep{George1991,Garcia2013} off a standard optically thick and
geometrically thin $\alpha$-disk \citep{Shakura1973}. Many ``bare''
and unabsorbed Seyfert 1 galaxies that are observed under shallow
inclination angles are supposed to allow a direct line-of-sight to the
inner disk and reflected radiation from this region, where strong
relativistic Doppler shifts, gravitational redshifts, and light bending
result in a significantly broadened iron line feature
\citep[e.g.,][and references therein]{Walton2013}.

The diagnostic power of broad iron lines for studying the accretion
physics in the presence of strong gravity has been used in a number of
models that are designed for fitting data in \texttt{XSPEC} and other
fitting packages. A model that allows to fit for the black hole spin
as a free parameter and to convolve a broad-band reflection continuum
with a relativistic kernel is provided with \texttt{relconv} by
\citet{Dauser2010}, who also include an overview of previous models
in this field. A number of studies using any of these models
consistently find larger disk emissivities closer to the black hole,
while the outer disk can be well described with a power-law emissivity
law of an $\alpha$-disk \citep[e.g.,][for MCG$-$6-30-15 as
well as a number of other AGN studied, e.g., by \citealt{Ponti2010},
\citealt{Brenneman2011}, \citealt{Wilkins2011}, \citealt{Dauser2012}
or
\citealt{Risaliti2013}]{Wilms2001,Fabian2002,Brenneman2006,Larsson2007}.
These steep inner emissivities strongly motivate a ``lamp-post''
geometry for the corona as opposed to a corona that sandwiches the
accretion disk \citep[][and references therein]{Svoboda2012}. In the
lamp-post geometry \citep{Martocchia1996}, the source of hard photons
that {irradiate} the accretion disk is located above the black
hole, on the {symmetry axis} of the system. A possible physical
realization would be that the corona is the base of a jet
\citep{Markoff2005,Wilkins2015,King2017}. The first model allowing to
directly fit the broad Fe~K\,$\alpha$ line in a self-consistent
lamp-post geometry has been published by \citet{Dauser2013}.  As a
further improvement, \citet{Garcia2014} link the reflected spectrum
\texttt{xillver} \citep{Garcia2013} from each point of an ionized disk
with the correct relativistic transfer function. Their model
\texttt{relxill\_lp} {is} the only model to date that allows
self-consistent fits of reflection {features under
  consideration of} a primary source at a certain height above the
disk in an angle-dependent way. A coverage of high signal-to-noise
(S/N) data above 10\,keV, as provided by \nustar, is important in such
models where the model parameters are very sensitive to changes in the
spectral shape \citep{Walton2016,Dauser2016}. A dedicated study of the
sample of bare AGN \citep{Walton2013} with \texttt{relxill\_lp} will
follow by Fink et al. (2017, in prep.).

Many continua of bare AGN, where most of the relativistically blurred
features have been detected, are free of strong and neutral absorption
but show signs of ionized warm absorption or outflows. Still,
Compton-thick absorption has been claimed to explain the broad
iron-line features of AGN \citep[e.g.,][]{Miller2008,Turner2009}.
\citet{Walton2010} provide strong arguments against this scenario and
in favor of inner-disk reflection.

The only intermediate-class Seyfert galaxy where a variable cold and
clumpy absorber
\citep{Risaliti2005,Risaliti2007,Risaliti2009a,Maiolino2010} has been
observed in conjunction with clear evidence for relativistic
reflection is NGC~1365
\citep{Risaliti2009b,Risaliti2013,Brenneman2013,Walton2013,Walton2014}. For
this source, \citet{Risaliti2009b} are able to disentangle a
partial-covering absorber with low covering fraction from similarly
broad spectral features of blurred reflection by making use of the
variability of the absorber. Mrk~766 may be an additional example,
however, the detection of a relativistically broadened iron line is
not yet clear \citep{Miller2007,Patrick2012}.

In Seyfert galaxies, variability in line-of-sight absorption across a
wide range of timescales has been observed.  One common
interpretation of such variability is the passage of discrete clouds
across the line-of-sight \citep{Nenkova2008a,Nenkova2008b}.  Clouds in
Cen~A \citep{Rivers2011a} and NGC~3227 \citep{Lamer2003} are observed
via centrally peaked column-density absorption profiles on
timescales of weeks to months and are inferred to reside in
the outer broad line region (BLR) or inner dusty torus.
Meanwhile, much shorter absorption events ($\lesssim1-3$ d) have been
detected, for example, in Mrk~766 \citep{Risaliti2011}, Fairall~9
\citep{Lohfink2012}, NGC~1365 \citep{Risaliti2007,Risaliti2009a},
Swift~J2127.4+5654 \citep{Sanfrutos2013}, or NGC~3227
\citep{Beuchert2015}. These patterns are consistent with transiting
clumps at the distance of the BLR.

NGC 4151 is a close-by ($z=0.003319$; \citealt{Vaucouleurs1991}) and
well-studied Seyfert 1.5 galaxy. Complex and variable line-of-sight
absorption has been reported from X-ray spectra for over three decades
\citep[e.g.,][]{Holt1980,Yaqoob1989,Fiore1990}. The absorption has
been modeled in a variety of ways across various X-ray missions;
successful models have typically incorporated combinations of
approximately two absorbers, for example, sometimes cold and/or warm
components. Having at least one partial-covering (covering fraction
typically $\sim 30$--70\%) component  is
{common}. Variability in absorption structure has been observed across a
wide range of timescales, but positively identifying the responsible
component(s) remains difficult
\citep{deRosa2007,Puccetti2007,Wang2010}.

\citet{Keck2015} have recently provided solid evidence for
relativistic reflection off the inner disk in NGC~4151 with additional
signs for absorption variability. The goal of this paper is to revisit
the spectral modeling of NGC~4151 and to pursue a physically-motivated
model of its broad-band X-ray spectrum and spectral variability. We
consider multiple datasets with high count statistics. We use the most
updated relativistic reflection model code, \textsc{relxillCp\_lp},
and remain mindful of potential degeneracies between relativistic
reflection and complex absorption \citep[e.g.,][for
NGC~4151]{Keck2015}. The choice of this model is strongly motivated by
the detection of non-relativistic radio jets at a velocity of
$\sim$$0.05\,c$ \citep{Wilson1982,Mundell2003,Ulvestad2005} and by
related Comptonization models in the jet-base of microquasars
\citep{Markoff2004,Markoff2005}. In Sect.~\ref{sec:obs} we provide an
overview of the observations we consider for the data analysis
presented in Sect.~\ref{sec:spectra}. For the analysis, we carefully
motivate a baseline model, which we then apply to all observations. We
also investigate the inherent spectral variability as well as the
sensitive parameters of the relativistic reflection components. The
soft X-rays are separately investigated. The results and implications
are discussed in Sect.~\ref{sec:discussion} and concluded in
Sect.~\ref{sec:conclusions}.

\section{Observations and data reduction}
\label{sec:obs}
\begin{table}
  \caption{\xmm\ and \suzaku\ observations in 2011/2012 with screened
    exposure times. Listed are the satellite, the observation ID of
    the observation, the date when the observation started, the
    exposure time after screening, and the number of counts detected.
    The check-symbol ($\surd$) denotes observations that are considered
    for the data analysis. Observations labeled with a cross
    (\ding{55}) are excluded. The observation Nu$_\mathrm{Suz}$ only
    contains data that are fully simultaneous to Suz~3. }
\label{tab:suzxmm_obs}
  \centering
  \scriptsize
  \resizebox{\columnwidth}{!}{
  \begin{tabular}[ht]{llllllllllllll}
    \hline\hline 
    abbrv. &obsid     & det   & time & exp [ks] & cnts [$\times 10^{4}$] &  \\
    \hline
    (XMM)   & 657840101 & pn & 2011-05-11 &   &     & \ding{55} \\
    &  & MOS 1 &           &  &  & \ding{55}  \\
    &  & MOS 2 &           &  &  & \ding{55}  \\
    XMM~1 & 657840201 & pn & 2011-06-12 & 2.3 &  1.4   & $\surd$ \\
    &  & MOS 1 &           & 0.2 & 0.07 & \ding{55}  \\
    &  & MOS 2 &           & 0.1 & 0.04 & \ding{55}  \\
    XMM~2 & 657840301 & pn & 2011-11-25 & 5.7 &  2.8   & $\surd$ \\
    &  & MOS 1 &           & 6.9 & 1.3  & $\surd$  \\
    &  & MOS 2 &           & 7.1 & 1.3 & $\surd$  \\
    XMM~3 & 657840401 & pn & 2011-12-09 & 6.6 &  3.0   & $\surd$ \\
    &  & MOS 1 &           & 8.8 & 1.5 & $\surd$  \\
    &  & MOS 2 &           & 9.0 & 1.5 & $\surd$  \\   
    XMM~4 & 679780101 & pn & 2012-05-13 & 6.3 & 2.6    & $\surd$ \\
    &  & MOS 1 &           & 8.6 & 1.3 & $\surd$  \\
    &  & MOS 2 &           & 8.7 & 1.3 & $\surd$  \\
    XMM~5 & 679780201 & pn & 2012-06-10 & 8.7 &  1.6   & $\surd$ \\
    &  & MOS 1 &           & 12.5 & 0.9 & $\surd$  \\
    &  & MOS 2 &           & 12.5 & 0.9 & $\surd$  \\
    XMM~6 & 679780301 & pn & 2012-11-14 & 3.8 & 2.4    & $\surd$ \\
    &  & MOS 1 &           & 1.8 & 0.2 & $\surd$  \\
    &  & MOS 2 &           & 2.0 & 0.3 & $\surd$  \\
    XMM~7 & 679780401 & pn & 2012-12-10 & 6.6 &  2.7   & $\surd$ \\
    &  & MOS 1 &           & 9.5 & 0.5 & $\surd$  \\
    &  & MOS 2 &           & 9.6 & 1.4 & $\surd$  \\
    (XMM) & 679780501 & pn & 2012-12-10 &  &     &\ding{55} \\
    &  & MOS 1 &           &  &  &\ding{55}  \\
    &  & MOS 2 &           &  &  & \ding{55}  \\    
    Suz~1 & 906006010 & XIS 0 & 2011-11-17 & 61.7 & 18.5  & $\surd$  \\
    & & XIS 1 &                  & 61.7& 17.7 &$\surd$   \\
    & & XIS 3 &                   &  61.7     &19.0  &$\surd$   \\
    & & HXD   &                   &  54.6      & 8.1 & $\surd$  \\
    Suz~2 & 906006020 & XIS 0 & 2011-12-18 & 55.1 & 21.1    & $\surd$ \\
    &  & XIS 1 &            &55.1 & 21.3 & $\surd$  \\
    &  & XIS 3 &            &55.1 & 23.4 & $\surd$ \\
    &  & HXD   &            & 32.1 & 7.3 & $\surd$   \\
    Suz~3 & 707024010 & XIS 0 & 2012-11-11 & 150.2 &   32.5   & $\surd$  \\
    &  & XIS 1 &            & 150.2& 32.8 & $\surd$ \\
    &  & XIS 3 &            & 150.2& 35.2 & $\surd$  \\
    &  & HXD   &           &  139.9 & 18.9 & $\surd$  \\
    Nu$_\mathrm{Suz}$ & 60001111002/3/5 & FPM~A & 2012-11-12/14 & 106.1 & 71.3 & $\surd$ \\
    &  & FPM~B & & 106.2 & 67.4 & $\surd$ \\
    \hline
  \end{tabular}
}
\end{table}

The observing log in Table~\ref{tab:suzxmm_obs} lists all examined
\xmm, \suzaku, and \nustar\ observations. They divide into the
long-look \suzaku\ observation Suz~3 with the simultaneous long-look
\nustar\ observation Nu$_\mathrm{Suz}$. These observations are
complemented with two additional \suzaku\ observations of significant
exposure, Suz~1 and Suz~2, from roughly one and two years before the
joint \suzaku/\nustar\ campaign. We also add a number of shorter \xmm\
observations in order to probe variability on timescales from months
to years.

\subsection{\suzaku}
In the analysis of the \suzaku observations we make use of data from the
X-ray Imaging Spectrometer \citep[XIS;][]{Koyama2007} and the Hard
X-ray Detector \citep[HXD;][]{Takahashi2007}. In particular, we use
the front (XIS~0,3) and back-illuminated (XIS1) chips in the
$3\times3$ and $5\times5$ editing modes and reprocess the unfiltered
event lists by applying standard procedures for event file screening
and attitude correction using \textsc{aeattcor2}. We use the
calibration releases 2015-10-05 for Suz~1 and Suz~2 and the later
version from 2016-02-04 for Suz~3. A previous investigation of Suz~3
by \citet{Keck2015} has shown discrepancies between the unfolded
spectra among the different XIS detectors below 2.5\,keV. With the
calibration from 2016-02-04, this effect seems to be reduced. We
therefore consider data in the full range between 0.6--10\,keV. We
find that both releases result in identical event files. For
\suzaku/HXD, we use the latest calibration release from 2011-09-15.
Spectra of both modes and all detectors are extracted from circular
regions of $\sim$90\arcsec\ radius, centered on the point source. We
exclude pixels above a threshold of 4\% pile-up as estimated with the
tool \textsc{pileest}. The resulting spectra of the $3\times 3$ and
$5\times 5$ modes are merged using \textsc{phaadd} for each XIS. In
order to guarantee sufficient statistics, all spectra are binned to a
minimal S/N of 10 but at least 11 channels per bin
in presence of spectral lines and at least 20 channels per bin for the
continuum. The size of each bin is larger than the resolution of the
response grid at
$\sim$6\,keV\footnote{\url{https://heasarc.gsfc.nasa.gov/docs/astroe/prop_tools/suzaku_td/node10.html}}.
Calibration uncertainties are prominent around the Si and Au K edges
at the energies of 1.8\,keV and $\sim$2.2\,keV, leading us to exclude
counts from the energy intervals 1.72--1.88\,keV and 2.19--2.37\,keV.

\begin{figure}
  \resizebox{\hsize}{!}{\includegraphics{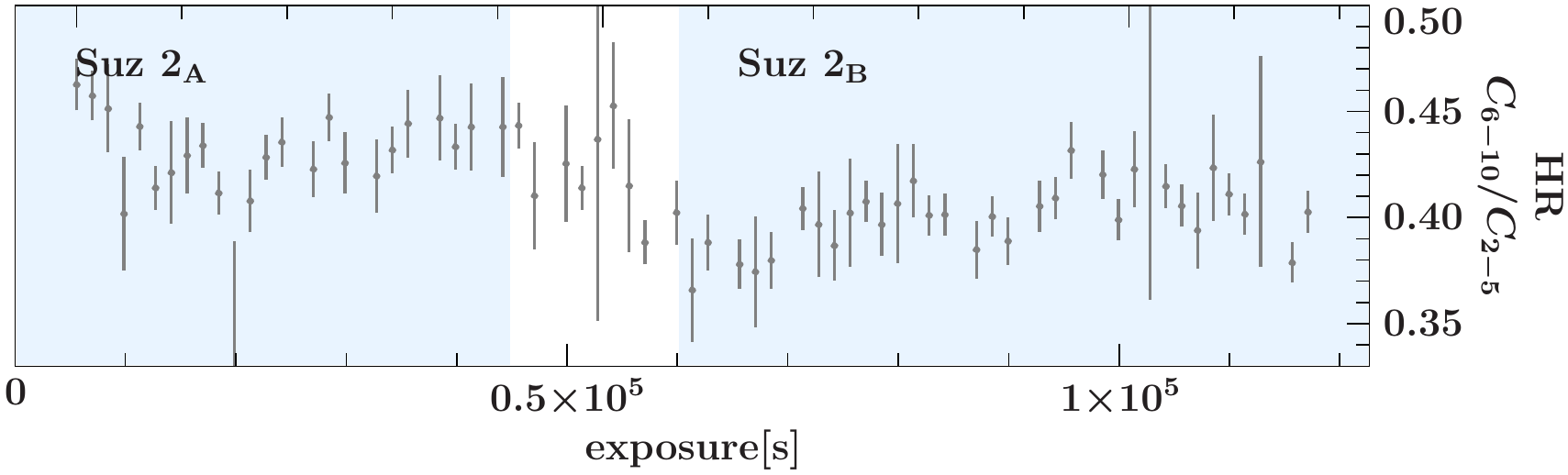}}
  \resizebox{\hsize}{!}{\includegraphics{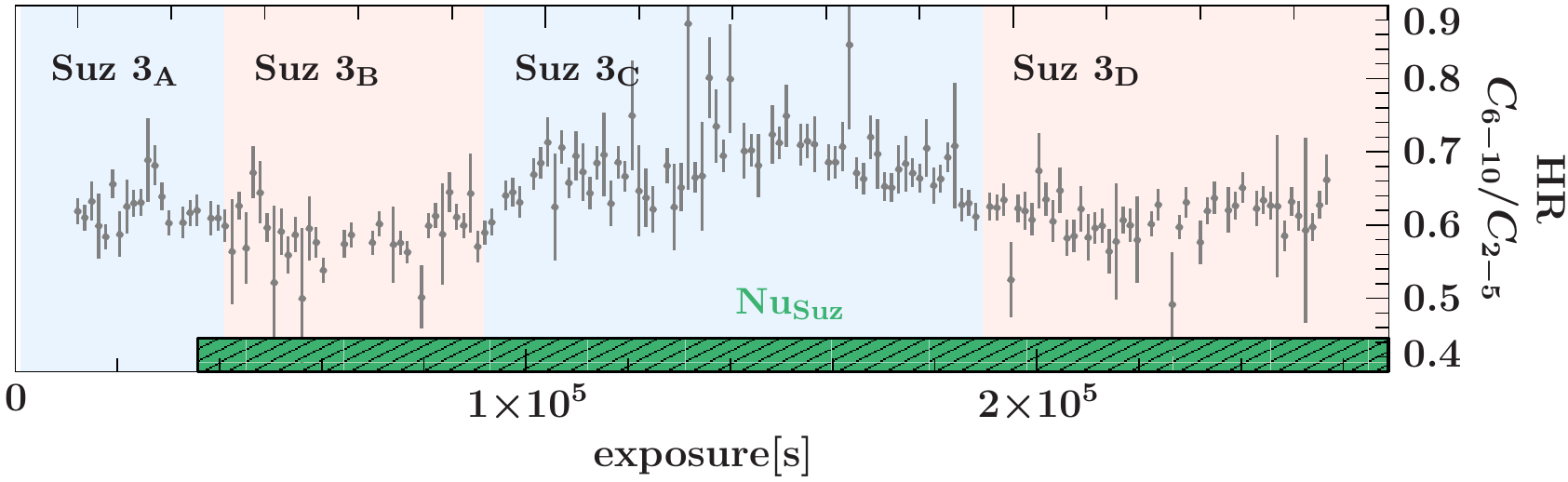}}
  \caption{Hardness ratio of the 55\,ks and 150\,ks \suzaku\
    observations Suz~2 (\textit{top}) and Suz~3 (\textit{bottom}) given by the ratio of
    count-rate light curves extracted between 6--10\,keV and
    2--5\,keV. The shaded regions divide the observation in different
    parts for a time-resolved spectroscopy and are derived using a
    Bayesian block analysis. The green-shaded and striped region in
    the bottom panel indicates the relative observation time of
    Nu$_\mathrm{Suz}$.}
  \label{fig:gti_suz_hr}
\end{figure}

Figure~\ref{fig:gti_suz_hr} shows the hardness ratio evolution for the
55\,ks and 150\,ks observations Suz~2 and Suz~3, respectively,
revealing significant variability.
We use the Bayesian block analysis \citep{Scargle1998,Scargle2013} to
divide both observations Suz~2 and Suz~3 into single sub-spectra
(Suz~2$_\mathrm{A}$, Suz~2$_\mathrm{B}$ and
Suz~3$_\mathrm{A}$--Suz~3$_\mathrm{D}$) that are highlighted in color
in the figure. This method uses Bayesian statistics to identify time
intervals that are, given the Poissonian likelihood, compatible with
constant hardness ratio. We extract observations from these blocks
with the same procedures as described above. The exposures divide into
25.0, 25.6, 55.9, and 43.7\,ks for the observations
Suz~3$_\mathrm{A}$--Suz~3$_\mathrm{D}$ resulting in $27.3\times
10^{4}$, $19.0\times 10^{4}$, $33.3\times 10^{4}$ , and $30.9\times
10^{4}$ counts for the combination (XIS~0+XIS~1+XIS~3).  Similarly, we
find 23.0 and 32.1\,ks for the observations Suz~2$_\mathrm{A}$ and
Suz~2$_\mathrm{B}$ with $23.1\times 10^{4}$ and $37.2\times 10^{4}$
counts.

The non-imaging HXD-PIN data are extracted for the whole $34\arcmin
\times 34\arcmin$ field of view.  The HXD data are binned to a minimum
S/N of 40, and a S/N of 20 for resolved sub-spectra. Where
simultaneous \nustar\ data are present, cross-calibration constants
are fitted relative to the \nustar\ focal plane module A
  (FPM~A) and compared to those found by \citet{Madsen2015}.
Otherwise spectra are normalized relative to
XIS0\footnote{http://heasarc.gsfc.nasa.gov/docs/suzaku/analysis/abc/}.

\subsection{\nustar}
\nustar\ \citep{Harrison2013} is the first instrument to focus hard
X-rays. X-rays are focused on the two focal-plane modules FPM~A and
FPM~B. We extract data from within 3--78\,keV with the standard
\nustar\ Data Analysis Software \texttt{NuSTARDAS-v.1.5.1}, which is
part of \texttt{HEASOFT-v.6.18}. Due to the high flux of NGC~4151, we
extract source counts from within a relatively large region of
90\arcsec\ radius on both chips FPM~A and FPM~B, and background counts
from a region of the same size located $\sim$340\arcsec\ off-source
but close enough not to introduce much bias due to the spatial
dependence of the background \citep{Wik2014}.  We explicitly extract
\nustar\ data from a time interval that is fully simultaneous to Suz~3
(Nu$_\mathrm{Suz}$) and from multiple intervals that are subsets of
Suz~3$_\mathrm{A}$--Suz~3$_\mathrm{D}$
(Nu$_\mathrm{Suz,A}$--Nu$_\mathrm{Suz,D}$) with exposures of 106\,ks,
2.5\,ks, 25.8\,ks, 50.2\,ks, and 27.6\,ks, respectively.  This
translates to a range of 3--$62\times 10^{4}$ counts for the
observations Nu$_\mathrm{Suz,A}$ to Nu$_\mathrm{Suz,D}$.  Spectra are
binned to a minimum S/N of 100 for the integrated spectrum and to 20
for the four individual spectra, which only leaves data below
50\,keV. Due to irregularities in the cross-calibration between \xmm,
\suzaku,\ and \nustar,\ only data taken above 5\,keV are considered for
\nustar.

\subsection{\xmm}
There exist nine observations of NGC~4151 with the EPIC camera
\citep{Strueder2001, Turner2001} between 2011-05-11 and 2012-12-10. In
addition, the \xmm\ observation with the ID 679780501 has to be
excluded due to poor data quality. All observations are in small-window
mode. We follow the standard procedure to extract data of all
detectors (EPIC-pn, EPIC-MOS1, EPIC-MOS2, and RGS) using the
\texttt{SAS~v.14} and the most recent calibration files. After
creating calibrated event lists with filtered hot and bad pixels,
events between 10 and 12\,keV are screened for particle flaring with a
threshold of $0.4\,\mathrm{cnts}\,\mathrm{s}^{-1}$. We extract all
counts within a maximum possible radius of 40--43\arcsec\ for EPIC-pn,
and within $\sim$120\arcsec\ for EPIC-MOS. While the region size is
physically limited by the chip border in the case of EPIC-pn, we are
able to extract counts from nearly 100\% of the encircled energy
fraction of the on-axis PSF for EPIC-MOS. Background counts are
extracted from an off-source spot on the chip within 45\arcsec\ and
89\arcsec\ for EPIC-pn and EPIC-MOS, respectively. We detect
significant pileup in all three cameras. In the case of the MOS, we exclude
the central pixels within 20\arcsec\ of the source position in all
cases. For the EPIC-pn, we exclude data from the inner 15\arcsec\ for
all observations, except for \texttt{0679780201}, where 17.5\arcsec\
needed to be excluded due to the higher count-rate of
8.7\,cnts\,s$^{-1}$. The EPIC-pn data are binned by a factor of 2
between 0.5--1.0\,keV, 4 between 1--5\,keV, 6 between 5--8\,keV, and
10 above that. The EPIC-MOS data are binned to a minimum S/N of 10,
with additional geometrical binning of 3, 5, and 12
channels/bin in the 0.5--1.0, 1.0--3.0, and 3.0--10\,keV
bands, respectively. This choice guarantees at least
  20--25\,cnts\,bin$^{-1}$ and provides an optimal trade-off between a
  decently binned continuum and sufficient data bins around line
  features in the spectrum. The first \xmm\ observation, XMM~1, will
not be further used due to strong particle flaring. For XMM~2, we
exclude the EPIC-MOS data due to the small amount of net-exposure
after filtering the event-files.

From the RGS \citep{denHerder2000} data, we extract both
orders with the task \texttt{rgsproc} and combine the spectra
  of the detectors RGS\,1 and RGS\,2 for each order with the task
\texttt{rgscombine}. When fitting the RGS data, we simultaneously
include both orders and all considered observations. We choose
  a geometrical binning of a factor of 3 for individual spectra to
  limit the oversampling of the theoretical RGS energy resolution
  as suggested by \citet{Kaastra2016}. Given the lack of sufficient
  counts per bin, we choose Cash-statistics for the further data
analysis.  Due to the low effective area at short wavelengths, we only
consider data below 1.3\,keV (9.5\AA).

\section{X-Ray spectral analysis}
\label{sec:spectra}
In the following, we examine all \xmm, \suzaku\ and \nustar\
observations. These observations were taken over a period of more than
one year. We provide a detailed investigation of the spectral
components as well as their variability. We apply the Galactic column
of $N_\mathrm{H,Gal}=2.3\times 10^{20}\,\mathrm{cm}^{-2}$
\citep{Kaberla2005} in all cases. In
Fig.~\ref{fig:ngc4151_all_xillver}, we show the spectra of all
observations in the top panel. The data imply a lack of obvious
variability of the soft X-rays below $\sim$1\,keV. The source is
moderately variable above 6\,keV within the range of 1--$1.5\times
10^{-3}\,\mathrm{Photons}\,\mathrm{cm}^{-2}\,\mathrm{s}^{-1}\,\mathrm{keV}^{-1}$
at 10\,keV with XMM~5 catching the source in an exceptionally low flux
state. Strong spectral variability, in contrast, is apparent for the
range between $\sim$1 and 6\,keV, both in spectral shape and
normalization. This argues strongly against a physical partial coverer
to explain the continuum including the soft emission below
$\sim$1\,keV \citep[see earlier work
by][]{Holt1980,Perola1986,Fiore1990} but in favor of diffuse emission
from optically thin and ionized gas on larger spatial scales
\citep[e.g.,][and references therein]{Wang2011c}.

\subsection{Motivation for a blurred-reflection component}
\label{subsec:motivation_relrefl}

\begin{figure}
\resizebox{\hsize}{!}{\includegraphics{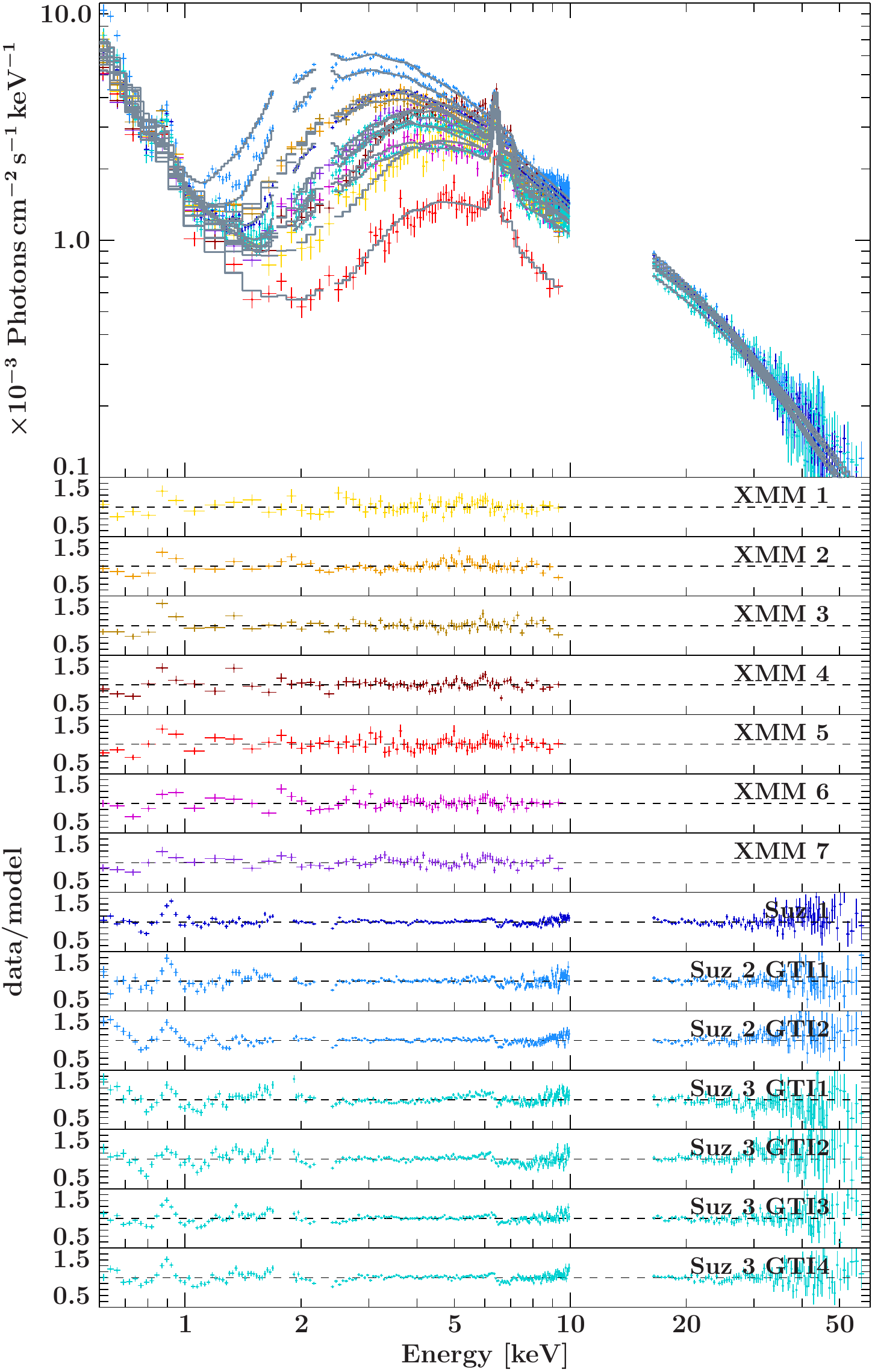}}
\caption{Fit to the \xmm\ and \suzaku\ data. The model consists of a
    power law that is partially covered by near-neutral intrinsic
    material as well as fully covered by Galactic foreground gas. A
    \texttt{xillver}-component of ionized, unblurred reflection is
    used to model the narrow iron line and potential soft-line
    emission.}
\label{fig:ngc4151_all_xillver}
\end{figure}

\begin{figure*}
\includegraphics[width=17cm]{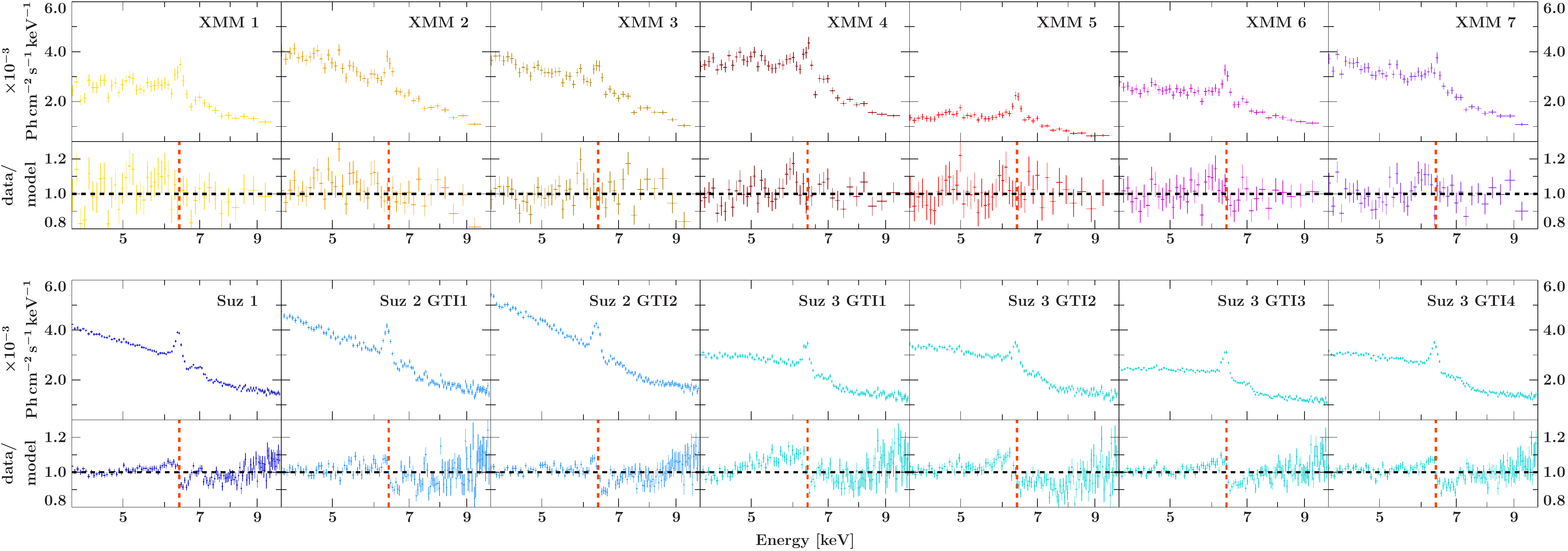}
\caption{Zoom-in into the iron-line region of the complete set of
    observations of \xmm\ and \suzaku\ with the same model applied as
    shown in Fig.~\ref{fig:ngc4151_all_xillver}. We mark the centroid
    energy of the Fe~K$\alpha$ line with a red, dashed line in the
    residual panels.}
\label{fig:ngc4151_all_xillver_ironline}
\end{figure*}

As a first approach to model the X-ray data, we simultaneously fit all
\xmm\ and \suzaku\ spectra with an incident and reflected power law
\citep[\texttt{xillver};][reflection off an optically thick,
geometrically thin, ionized accretion disk]{Garcia2013} that is
absorbed by near-neutral intrinsic material (\texttt{zxipcf}) and
Galactic foreground gas (\texttt{tbnew}). We fit the diffuse emission
below 1\,keV phenomenologically with a partial-covering version of
\texttt{zxipcf} with a covering fraction of $\sim$98\%. We emphasize
that the partial coverer has no physical meaning in this case and
reflects the combination of a fully absorbed with an attenuated power
law, both of identical slope. A physical interpretation of the latter
would be nuclear emission scattered off distant and large-scale gas.
Figure~\ref{fig:ngc4151_all_xillver} shows the spectra with
corresponding fits. The modeling confirms a strongly changing column
density ranging between 2.6 and $14.0\times 10^{22}\,\mathrm{cm}^{-2}$ as
well as a variable power-law flux. The ionization states of the
reflecting material are found to lie between $\log \xi \sim -3$ and $ -1.1$.

For simplicity, we freeze the cutoff energy of the incident continuum
to $E_\mathrm{cut}=300\,\mathrm{keV}$ and the inclination angle to
$i=30^\circ$. The Galactic column is considered as foreground
absorption in all spectral fits. The residuals indicate decent fits to
the continua of the individual observations and the ionized reflection
component accounts well for a narrow Fe~K$\alpha$ and Fe~K$\beta$ line
contribution but leaves line-like residuals below 2\,keV,
strengthening the notion of line-emitting plasma. Broad line-like
residuals between $\sim$5 and 6.4\,keV, and a hard excess above 8\,keV
are reminiscent of an additional component of blurred reflection. See
also Fig.~\ref{fig:ngc4151_all_xillver_ironline} for a zoom into these
features.

\subsection{Building on a recent investigation of the \suzaku\ data:
  application of an improved model for relativistic reflection}
\label{subsec:oldmodel}
\begin{figure}
  \resizebox{\hsize}{!}{\includegraphics{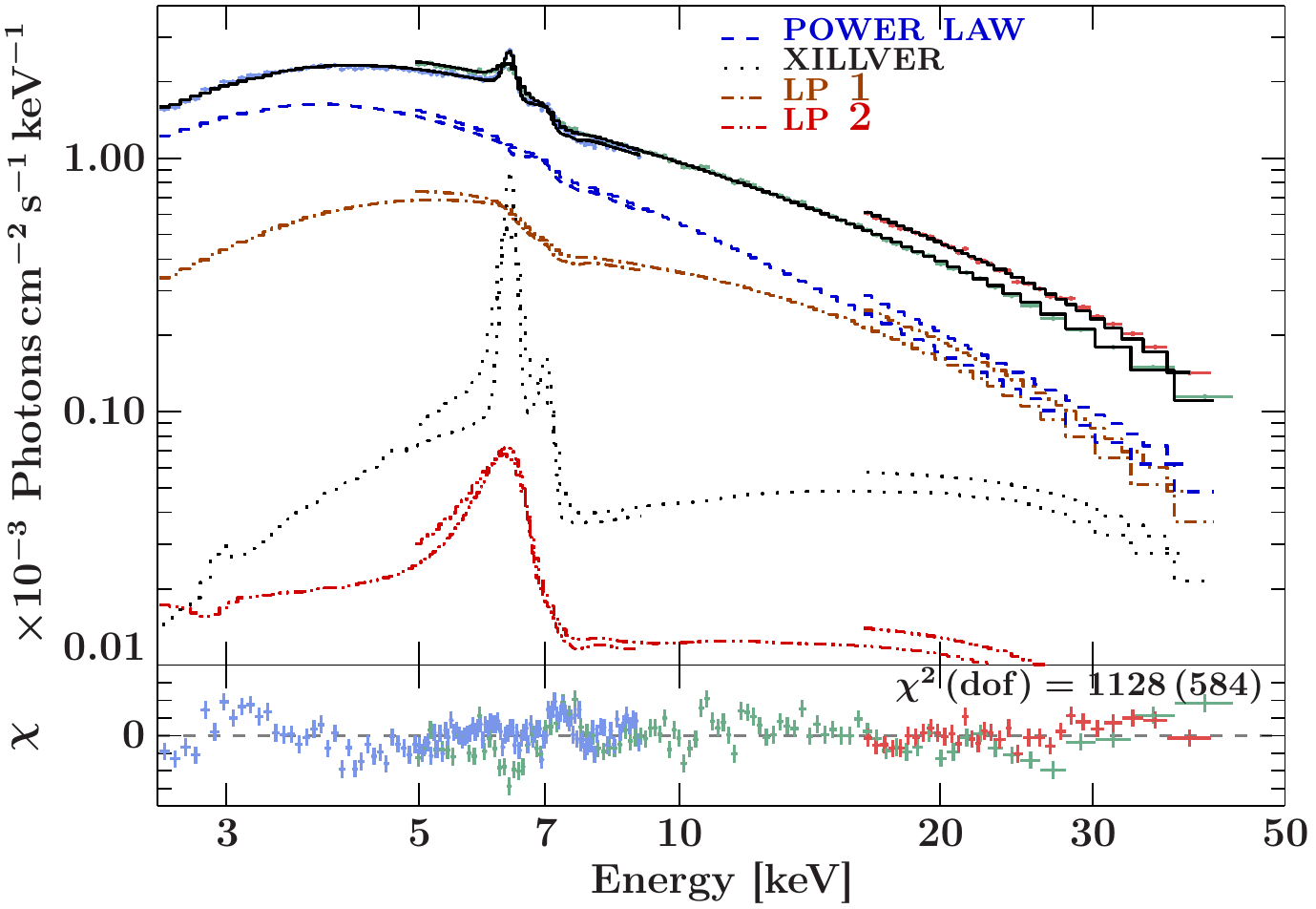}}
  \caption{ Application of the two-lamp-post model of \citet{Keck2015}
    to the Suz~3 and Nu$_\mathrm{Suz}$ data. All model components
    shown here are fully absorbed with the column reported by
    \citet{Keck2015}. The incident continuum is plotted in blue, the
    unblurred component in black and the blurred reflection components
    with a low and high primary source in brown and red,
    respectively. The residuals indicate the \suzaku/XIS, \suzaku/HXD
    and \nustar\ data in blue, green, and red, respectively.}
  \label{fig:relxill_keck}
\end{figure}

\citet{Keck2015} have presented a detailed study of the joint
\suzaku/\nustar campaign (Suz~3/$\mathrm{Nu}_\mathrm{Suz})$ for data
above 2.5\,keV. They discuss two possible models to explain the data:
(1) their best-fit model dominated by relativistic inner-disk
reflection and (2) an entirely absorption-dominated model.  The first
is given by a convolution of the unblurred and initially
angle-resolved reflection continuum \texttt{xillver} with the
relativistic code \texttt{relconv} \citep{Dauser2010} describing
reflection off the inner parts of an accretion disk.  The authors also
test for the more self-consistent model \texttt{relxill}
\citep{Garcia2013}, which links the relativistic transfer-function and
the angle-resolved disk-reflection spectrum \texttt{xillver} at each
point of the disk. Although \citet{Garcia2013} predict deviations of
up to 20\% for parameters of \texttt{relconv} and \texttt{relxill},
\citet{Keck2015} find a negligible statistical difference. They
attempt to fit for a lamp-post geometry with the convolution code
\texttt{relconv\_lp} \citep{Dauser2013} and require two distinct
lamp-post components at different heights ($\sim 1.3\,r_\mathrm{EH}$
and $\sim 14\,r_\mathrm{EH}$, where $r_\mathrm{EH}$ is the event
horizon). The authors, however, find significant S-shaped residuals
between 3--5\,keV and therefore reject this solution on statistical
grounds. The immediate aim of our study is to investigate this model
description using a lamp-post geometry that is physically motivated
not only by steep inner emissivities \citep{Svoboda2012} but also by
independent reverberation studies \citep[e.g.,][and references
therein]{Kara2013}. As a first improvement, we replace the convolution
model with the self-consistent and fully angle-resolved version
\texttt{relxill\_lp} and adopt the parameters found by
\citet{Keck2015}. In Fig.~\ref{fig:relxill_keck}, we show this model
evaluated for the data of Suz~3 and Nu$_\mathrm{Suz}$ in the same
energy range considered by \citet{Keck2015}. We find similar
statistics and residuals as compared to the convolution model. The
S-shaped residuals around the turnover of the absorber between
3 and 5\,keV imply a yet unmodeled partial coverer, which the authors
investigated as part of their independent model (2).

This model combines two partial coverers, one with a column of $\sim
6\times 10^{23}\,\mathrm{cm}^{-2}$ and low covering fraction
($\sim$40\%) and a second one with $\sim 1.3\times
10^{23}\,\mathrm{cm}^{-2}$ and a near-maximum covering fraction
($\sim$94\%). The model provides a decent description of the continuum
above 2.5\,keV, but leaves residuals reminiscent of a broad iron line
and is statistically less preferred. This demonstrates that modeling
the complex spectrum of NGC~4151 is not straightforward but combined
with degeneracies between these two solutions.

In the following we build a baseline model using the long-look 150\,ks
Suz~3 and simultaneous Nu$_\mathrm{Suz}$ observations. We extend the
work by \citet{Keck2015} by considering the entire energy range
covered by the instruments. In order to flatten the S-shaped residuals
in our broad-band X-ray continuum, we combine their models (1) and (2)
and apply the latest self-consistent relativistic reflection code
\texttt{relxillCp\_lp}\footnote{The model \texttt{relxillCp\_lp}
  extends on \texttt{relxill\_lp} \citep{Dauser2013} taking into
  account the Comptonization continuum \texttt{nthcomp} as primary
  continuum.}, which is, similar to the latest version of
\texttt{xillver}, calculated using the Comptonization continuum
\texttt{nthcomp} \citep{Zdziarski1996}. In the following, we fix the
seed photon temperature to 50\,eV, which is used to calculate the
\texttt{xillver} tables.

\subsection{A re-investigation of the 150\,ks simultaneous \suzaku\
  and \nustar\ observations with \texttt{relxillCp\_lp}}
\label{sec:suz3fit}
\subsubsection{Derivation of a baseline model} 
\label{sec:suz3fitbaseline}
\begin{figure}
  \resizebox{\hsize}{!}{\includegraphics{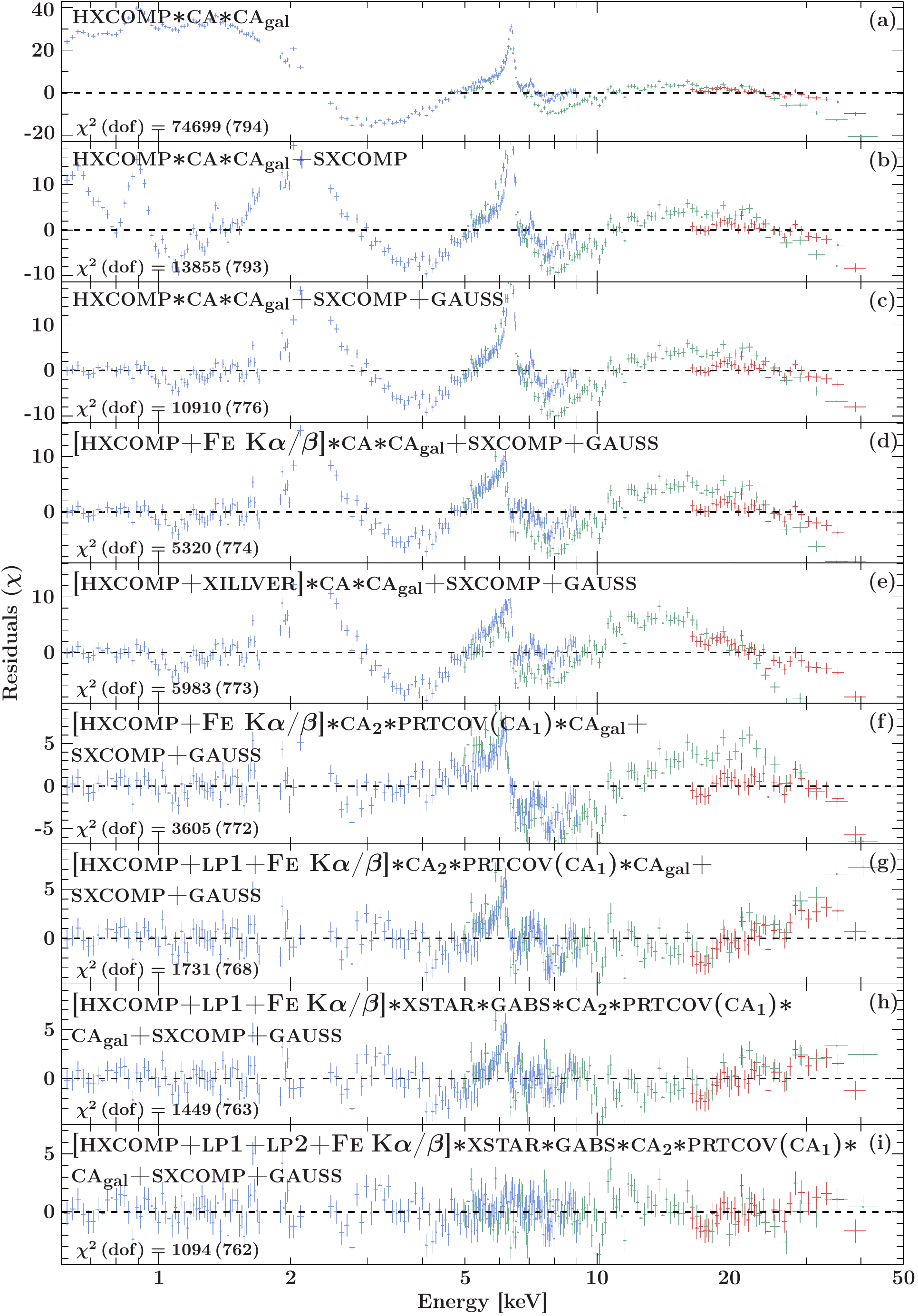}}
  \caption{Residuals ($\chi$) for a bottom-up approach towards the
    best-fit of Suz3. The statistics of each step are shown on the
    bottom left. \nustar\ residuals below 5\,keV are excluded from the
    plot due to deviations of the cross-calibration between \suzaku\
    and \nustar. The ranges of the residual axes are adapted in order
    to show the full dynamic range.}
  \label{fig:suz3_bottom_up}
\end{figure}
The residuals at each step are shown in Fig.~\ref{fig:suz3_bottom_up}.
First, we fit the data with the Comptonization continuum
\texttt{nthcomp}, hereafter referred to as HXCOMP. It provides a more
physically motivated primary continuum with an intrinsic cutoff as
opposed to a power law with external cutoff \citep{Garcia2015}. This
continuum is absorbed by fully-covering neutral gas (CA,
\texttt{tbnew\_simple\_z}). The residuals are shown in
Fig.~\ref{fig:suz3_bottom_up}a and in particular indicate an unmodeled
soft continuum.

In previous studies using \chandra\ data, the soft continuum was
accounted for with a bremsstrahlung component
\citep{Ogle2000,Wang2011c}. The latter authors, however, emphasize
their lack of a physical motivation for this component. We therefore
adopt a simple scenario, in which the nuclear Comptonized continuum is
scattered off distant and large-scale gas and model this component
with an unabsorbed soft Comptonization continuum (SXCOMP,
Fig.~\ref{fig:suz3_bottom_up}b). Here, we adopt the same parameters as for
the HXCOMP but leave the normalization free to vary. This way of
modeling introduces a minimal set of additional degrees of
freedom. Also, at CCD energy resolution and given the blend of
emission lines present (see below), we cannot constrain the exact form
of the soft continuum. Fits using Comptonization or
bremsstrahlung continua yield statistically identical fits.

On top of the soft continuum, a number of lines appear in the
residuals, reminiscent of emission from the ionized large-scale gas
component that has been extensively studied by
\citet{Wang2011c}, for example, using \chandra\ and physical emission
  codes. We focus mainly on the nuclear properties of the
  X-ray spectrum and instead fit this emission with a
phenomenological blend of Gaussians
(Fig.~\ref{fig:suz3_bottom_up}c). The centroid energies are adopted
from those lines that are significantly detected in gratings data of
\xmm/RGS \citep{Schurch2004} and \chandra/LETG \citep[][see also
\citealt{Vainshtein1978}]{Ogle2000}, or as line-blends by
\chandra/ACIS \citep{Wang2011c}. Table~\ref{tab:gaussians} lists the
parameters of all fitted lines with their ion identification, line
flux and centroid energies. For deriving the final photon line fluxes
and their uncertainties, we freeze all continuum parameters
including the SXCOMP normalization to the best-fit value found in the
final step. This also reduces degeneracies between the line-blend and
the continuum. The remaining residuals show a yet unmodeled swing in
the continuum around 2\,keV, strong and broad emission forming an
iron-line complex at approximately 6\,keV and extra curvature above
10\,keV.

Narrow emission components at the centroid energies of the
Fe~K$\alpha\,(\beta)$ lines at $\sim$6.4\,(7.1)\,keV can be modeled
with either two narrow Gaussians
\citep{Warwick1989,Zdziarski2002,Schurch2003,Wang2010} with a frozen
flux-ratio of 12\% (Fig.~\ref{fig:suz3_bottom_up}d) or a component of
distant reflection (\texttt{xillver}) in
Fig.~\ref{fig:suz3_bottom_up}e.  Both options leave a broad emission
feature between 5\,keV and 6\,keV, as previously seen by
\citet{Wang1999}, as well as equal continuum residuals. We are
therefore not able to confirm a strong statistical need for a distant
reflection component. We continue to use two Gaussian components as
a phenomenological description of the iron line emission that likely
originates from Compton-thin gas. This way, we only add a minimal set
of additional degrees of freedom to the model.

The continuum results in S-shaped residuals below 5\,keV, similar to
what \citet{Keck2015} found.  We replace the fully covering absorber
(CA) with the combination of a partially covering absorber (CA$_{1}$)
and a fully covering absorber (CA$_{2}$), both neutral (see
Fig.~\ref{fig:suz3_bottom_up}f). CA$_{1}$ requires a covering fraction
of $\sim$40--50\%. A similar dual neutral absorber has frequently been
applied before \citep[e.g.,][and references
therein]{Wang2010,Keck2015}. Further tests reveal that the introduced
spectral curvature can not be reproduced with an ionized warm
absorber.

The broad pattern just below 6\,keV remains, even after flattening the
continuum below 5\,keV. It likely features the red wing of an
extremely blurred Fe~K$\alpha$ line \citep[e.g.,][]{Dauser2010}. We
attempt to account for this feature with the previously introduced
model \texttt{relxillCp\_lp}. We call this component LP$_1$. Its
height hits the lower limit at $\sim 1.1\,r_\mathrm{EH}$, which
corresponds to $1.2\,r_\mathrm{g}$ for the spin fixed at its maximum
value. Note that we fit the incident and reflected continua
independently at this point. Due to the lack of data above 50\,keV, we
are unable to constrain the cutoff energy (expressed via the electron
temperature in our model) and freeze it at
$kT_\mathrm{e}=399\,\mathrm{keV}$ or $E_\mathrm{cut}\approx
1000\,\mathrm{keV}$. We note, however, that \citet{Malizia2014} find
$E_\mathrm{cut}=196_{-32}^{+47}\,\mathrm{keV}$ with \integral, \swift,
and \xmm\ data between 2 and 100\,keV. We expect significant bias for
their measurement of the cutoff due to: the lack of soft X-rays
\citep{Garcia2015}; the use of a lower photon index of
$\Gamma\sim$1.63; and the background dominance of \integral\ as
opposed to the \nustar\ data used in this work. We emphasize that for
primary sources close to the black hole, the cutoff energy has to be
corrected for the gravitational redshift as outlined by
\citet{Niedzwiecki2016}.  The resulting residuals
(Fig.~\ref{fig:suz3_bottom_up}g) illustrate that the underlying broad
feature as well as parts of the Compton hump above 10\,keV are
successfully fitted by LP$_1$.

Dips around 7\,keV suggest an additional column of highly ionized
absorption. We can greatly improve the fit using a \texttt{XSTAR}
absorption component for dense, coronal gas with $\log \xi \sim
2.8$. Warm absorber components with similar ionization have previously
been found by \citet{Weaver1994b}, \citet{Schurch2002}, and
\citet{Keck2015}.  The \texttt{XSTAR} model removes line-like
residuals close to the centroid energies of
\ion{Fe}{xxv}~He\,$\alpha$, \ion{Fe}{xxvi}~Ly\,$\alpha,$ and
\ion{Fe}{xxv}~He\,$\beta$, at zero velocity offset relative to
systemic.  There remain narrow absorption-like residuals around
8\,keV; they can be modeled with a broad Gaussian absorption component
that has an energy centroid of $8.17^{+0.11}_{-0.09}\,\mathrm{keV}$
(rest frame).  We identify this feature as a blueshifted
\ion{Fe}{xxv}~He\,$\alpha$ or \ion{Fe}{xxvi}~Ly\,$\alpha$ line,
although the possibility of unmodeled contributions from
\ion{Fe}{xxvi}~Ly\,$\beta$ (8.25\,keV rest frame) associated with the
systemic warm absorber cannot be ruled out.  The implied velocities
are $0.22^{+0.02}_{-0.01}\,\mathrm{c}$ for He-like Fe or
$0.17^{+0.02}_{-0.01}\,\mathrm{c}$ for H-like Fe, suggesting an
ultra-fast outflow (UFO), as previously constrained with a highly
ionized \texttt{XSTAR} component \citep[see][for a physical modeling
of the UFO with \texttt{XSTAR}]{Tombesi2011,Tombesi2013ufo}. The
flattened residuals are shown in Fig.~\ref{fig:suz3_bottom_up}h.

An excess around $\sim$6\,keV implies a still broad iron line with a
much less smeared red wing. We therefore add a second lamp-post
component (LP$_2$) with a primary source at the larger height of $\sim
17\,r_\mathrm{EH}$, which also takes care of remaining excess
residuals above 20\,keV. While both components LP$_1$ and LP$_2$
already describe the entire Compton hump above 10\,keV, further tests
can exclude the need for a third, distant reflection component to
model also the narrow iron line. We find a best-fitting baseline model
with $\chi^{2}/\mathrm{dof}=1094/762$ and overall flat residuals in
Fig.~\ref{fig:suz3_bottom_up}i.

\subsubsection{The best-fit baseline model} 
\begin{figure}
    \resizebox{\hsize}{!}{\includegraphics{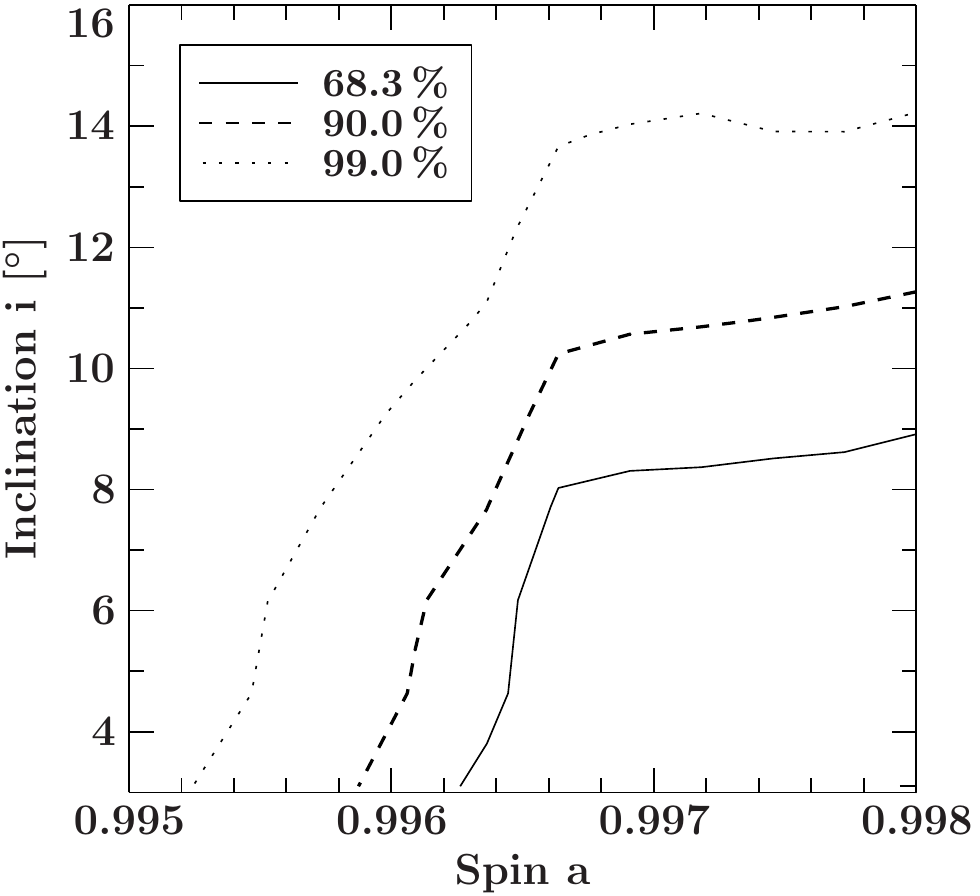}}
    \caption{Contours between the spin and inclination of both
      lamp-post reflection components for a fit with free spin. We
      show 68.3\%, 90\%, and 99\% confidence levels.}
  \label{fig:cont_a_i}
\end{figure}
\begin{figure}
    \resizebox{\hsize}{!}{\includegraphics{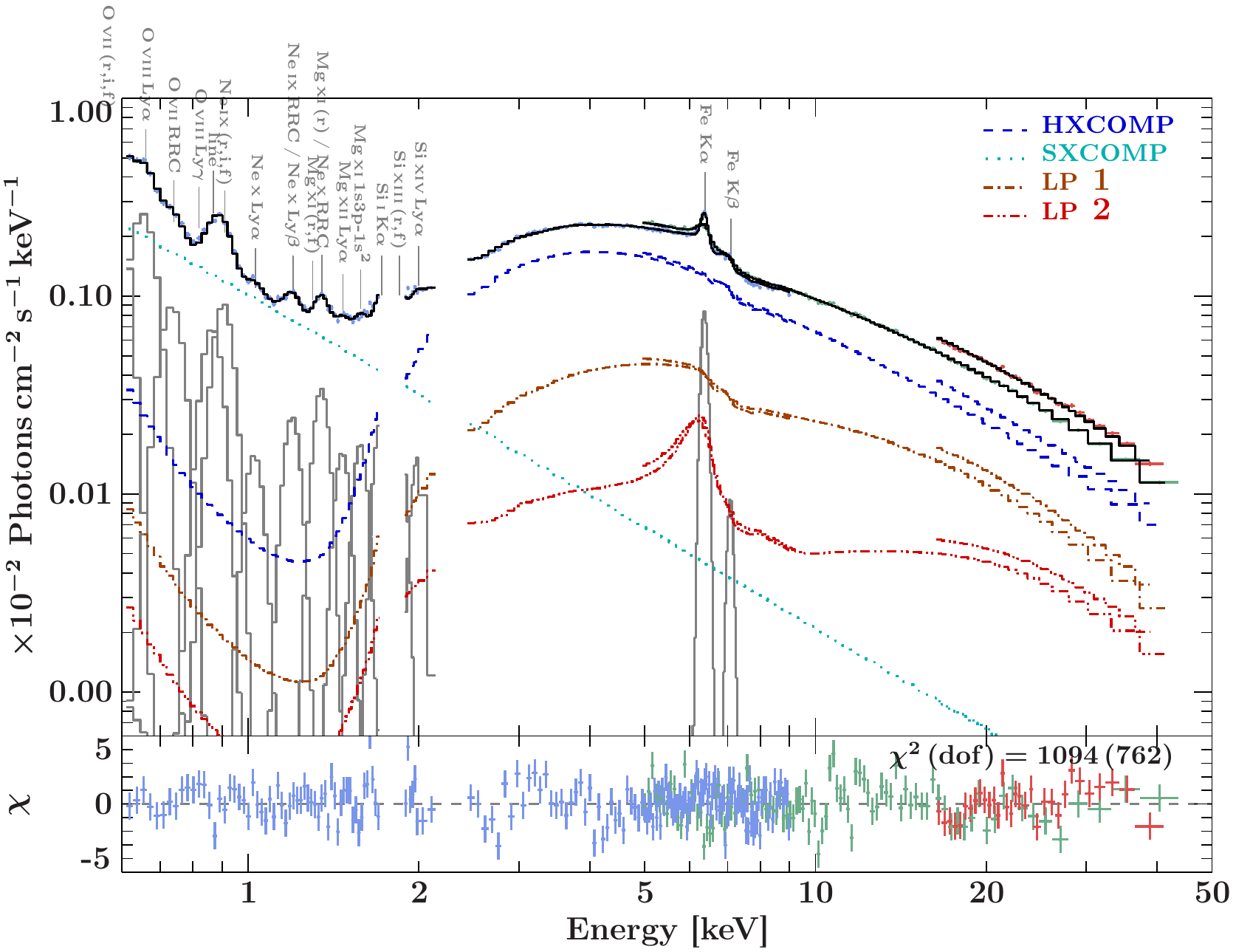}}
    \caption{Model components together with the data and residuals of
      the 150\,ks long observation Suz~3. The incident HXCOMP
      continuum (dashed blue line) and both lamp-post models (LP$_1$:
      dotted-dashed brown line, LP$_2$: double-dotted-dashed red line)
      are shown as absorbed components, the SXCOMP (dotted cyan line)
      as unabsorbed component. The soft emission lines (gray) required
      for the best-fit are labeled with their line identifiers and
      listed in Table~\ref{tab:gaussians}. The residuals of
      \suzaku/XIS, \suzaku/HXD, and \nustar/FPM, are drawn in
      blue, red, and green, respectively.}
  \label{fig:modcomp_suz3_avrg}
\end{figure}
We found a solid model for the broad continuum from the soft to the
hard X-rays, which, below, we apply to the remaining
observations. This baseline model combines a set of four complex
absorbers with a physical description of blurred reflection as part of
the self-consistent lamp-post geometry. We fit both LP components with
a tied slab-ionization and find a common value of $\log \xi \sim
2.8$. The fit is hitting the lower-limit for the inclination of
3.1$^\circ$ with 90\% uncertainties allowing values as high as
10$^\circ$ (see the contours between disk inclination and spin in
Fig.~\ref{fig:cont_a_i} where we unfroze the spin parameter). The wide
range originates in the $\cos i$-dependence of the model. This
inclination is consistent with the constraint of $\theta < 30^\circ$
that \citet{Cackett2014} found with reflection-component reverberation
mapping. We find no obvious correlation with the spin parameter and
fix the spin at its maximum value. The relatively flat turnover
between 3 and 6\,keV is well described by the dual neutral absorber
rather than a single full-covering absorber.  The column densities of
the two neutral absorbers are found to be $N_\mathrm{H,1}\sim 21\times
10^{22}\,\mathrm{cm}^{-2}$ with $f_\mathrm{cov}=0.46$ and
$N_\mathrm{H,2}\sim 9\times 10^{22}\,\mathrm{cm}^{-2}$ for the
full-covering absorber. The HXCOMP photon index is well constrained to
$\Gamma=1.72\pm0.01$ for the long-look observation and will therefore
be kept fixed for the remaining observations of lower count statistics
in order to reduce degeneracies within the complex model composite.
We list the best-fit parameters and their uncertainties in
Table~\ref{tab:bestfit} and show all model components in
Fig.~\ref{fig:modcomp_suz3_avrg}.
\begin{figure}
  \resizebox{\hsize}{!}{\includegraphics{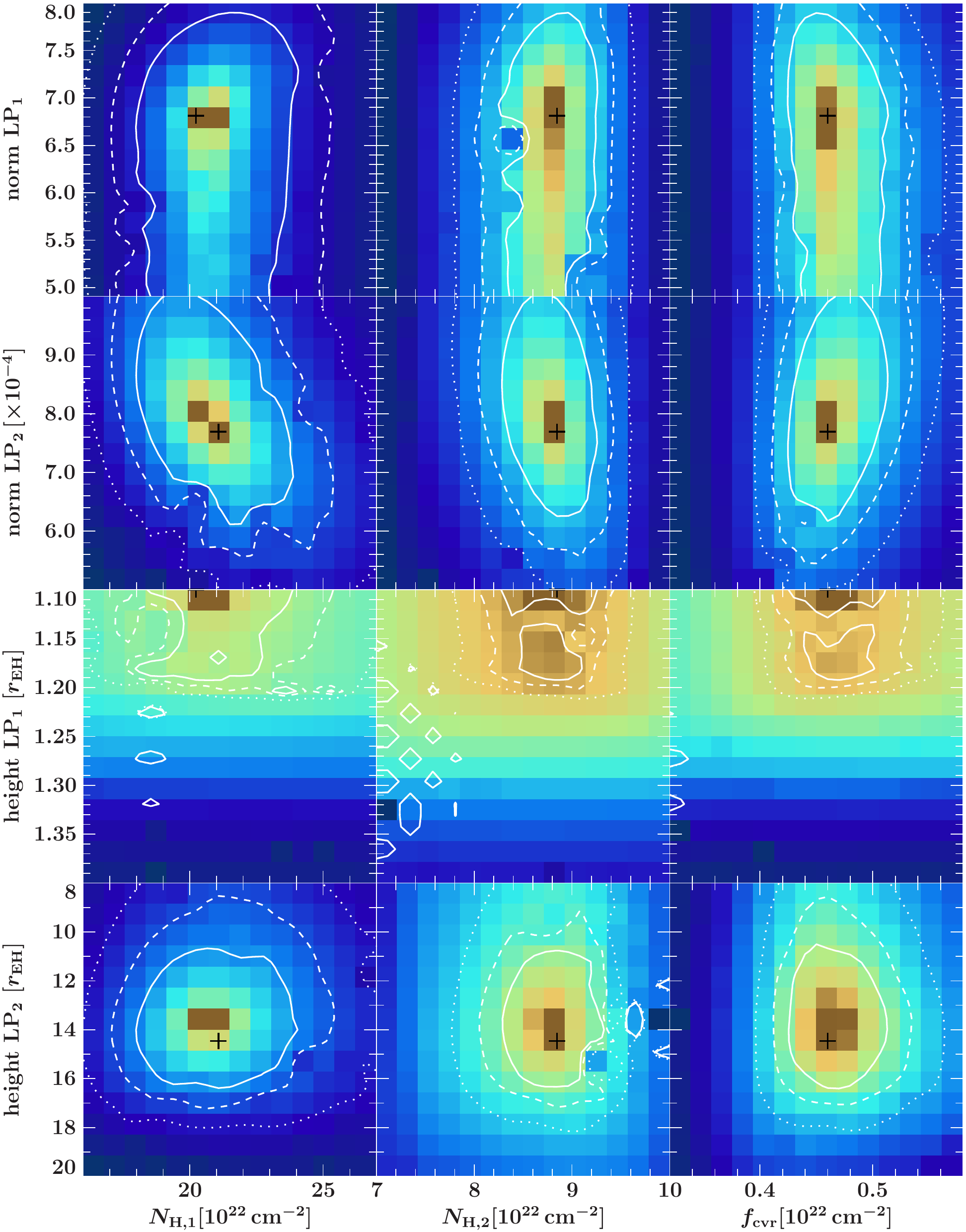}}
  \caption{$\Delta\chi^2$-maps relating the parameters of the cold
    absorbers CA$_1$ (partial covering) and CA$_2$ with those of the
    blurred reflection components LP$_1$ and LP$_2$. We show three
    contour levels of 68.27\%, 90\%, and 99\% (solid, dashed, and
    dotted lines).}
  \label{fig:suz3_lp_contours}
\end{figure}

\begin{table}
  \caption{Best-fit parameters of the blend of Gaussian lines for Suz~3. All Gaussian lines are fitted
    with zero width. Centroid energies marked with $\infty$ are adopted from
    \citet{Schurch2004}, \citet{Ogle2000}, and \citet{Vainshtein1978},
    line blends marked with $\dagger$ from the \chandra/ACIS
    study of \citet{Wang2011c}. Frozen parameters are denoted with an
    asterisk ($\ast$). When marked with the symbol $\pisces$, the line identification is uncertain and can be confused with an instrumental edge. The lines are consistent with the study on
    \xmm/RGS data in this work. Note
    that the \ion{O}{vii} line falls outside the sensitive energy range of \suzaku/XIS. For the possible line-blends
    \ion{Ne}{ix}~RRC / \ion{Mg}{xi}~r and \ion{Si}{xiii}~(r/f), we additionally
    fit for the centroid energies, because of unclear line identifications. Also, we
    require two narrow Gaussians with zero width at 1.58\,keV and
    1.72\,keV, which may be due to intrinsic \ion{Si}{i}~K$\alpha$ and
    \ion{Mg}{xi}~$1s3p-1s^{2}$ or due to calibration effects at the
    instrumental Al~K and Si~K edges.} \resizebox{\columnwidth}{!}{
\begin{tabular}[ht]{llll}
\hline\hline 
Line & E$^\ast$~[keV]~/~$\lambda^\ast$~[\AA] & flux~[Ph\,s$^{-1}$\,cm$^{-2}$] \\
\hline
\ion{O}{vii} (r,i,f)\,$^\dagger$ & 0.57 / 21.80 & $\left(1.1\pm0.3\right)\times10^{-3}$\\
\ion{O}{viii}~Ly\,$\alpha$\,$^\infty$ & 0.65 / 18.97 & $\left(2.67\pm0.19\right)\times10^{-4}$ \\
\ion{O}{vii} RRC\,$^\infty$ & 0.74 / 16.77 & $\left(9.4\pm1.0\right)\times10^{-5}$\\
\ion{O}{viii} RRC\,$^\infty$ & 0.87 / 14 & $\left(7.6^{+0.7}_{-1.0}\right)\times10^{-5}$\\
\ion{Ne}{ix} (r,i,f)\,$^\dagger$ & 0.91 / 13.55 & $\left(1.02\pm0.09\right)\times10^{-4}$\\
\ion{Ne}{x}~Ly\,$\alpha$\,$^\dagger$ & 1.03 / 12.04  & $\left(1.8\pm0.4\right)\times10^{-5}$\\
\ion{Ne}{ix}~RRC~/~\ion{Ne}{x}~Ly\,$\beta$\,$^\dagger$ & 1.2 / 10.33 & $\left(3.0\pm0.4\right)\times10^{-5}$\\
\ion{Ne}{x}~RRC~/~\ion{Mg}{xi} (r)\,$^\infty$ & $1.351^{+0.004}_{-0.005}$ / $9.18\pm 0.03$ & $\left(4.4\pm0.3\right)\times10^{-5}$\\
\ion{Mg}{vii}~Ly\,$\alpha$\,$^\infty$ & 1.47 / 8.43 & $\left(2.2\pm0.3\right)\times10^{-5}$ \\
\ion{Mg}{xi}~$1s3p-1s^{2}$\,$^\infty~\pisces$ & 1.58 / 7.85 & $\left(2.8\pm0.4\right)\times10^{-5}$\\
\ion{Si}{i}~K$\alpha$\,$^\infty~\pisces$ & $1.7200^{+0.0017}_{-0.0000}$ / $7.2084^{+0.0000}_{-0.0071}$ & $\left(3.6\pm0.4\right)\times10^{-5}$\\
\ion{Si}{xiii} (r,f)\,$^\infty$ & $1.8473^{+0.0008}_{-0.0013}$ / $6.712^{+0.005}_{-0.003}$ & $\left(4.4\pm0.4\right)\times10^{-5}$\\
\ion{Si}{xiv}~Ly\,$\alpha$\,$^\infty$ & 2.0 / 6.20 & $\left(2.3\pm0.4\right)\times10^{-5}$\\
\hline\hline 
Continuum & $\Gamma$ & norm [Ph\,keV$^{-1}$\,s$^{-1}$\,cm$^{-2}$] & \\
\hline
 & 1.72$^\ast$ & $\left(1.33\pm0.05\right)\times10^{-3}\,^\ast$ &  \\
\hline
\end{tabular}
}
\label{tab:gaussians}
\end{table}

\begin{table}
  \caption{Best-fit parameters of the baseline model for the
    simultaneous 150\,ks data of Suz~3 and Nu$_\mathrm{Suz}$. The
    black hole spin of both lamp-post components is set to its maximum
    value of $a=0.998$, while the radii of the inner and outer disk
    are kept at the default values of $1\,r_\mathrm{EH}$ and
    $499\,r_\mathrm{g}$, respectively. Parameters marked with the
    symbol $\dagger$ are tied amongst one another while those marked with
    an asterisk ($\ast$) are frozen. The reflection fraction is no
    free parameter here, as the incident continuum (\texttt{nthcomp})
    and both reflection continua (LP$_1$, LP$_2$) are fitted
    independently (\texttt{refl\_frac=-1}). The normalization of
    \texttt{nthcomp} is defined at unity for a norm of 1 at 1\,keV.
    The normalization of \texttt{xillver} and \texttt{relxill} is
    defined in the Appendix of \citet{Dauser2016}.} \centering\small
  \renewcommand{\arraystretch}{1.3}
\begin{tabular}[ht]{llll}
  \hline\hline 
  Model component & Parameter & Value\\
  \hline
  \multicolumn{3}{c}{$\chi^2$\,(dof) = 1094\,(762)}\\
  \hline
  Detconst & XIS\,0  & $0.998\pm0.005$\\
  & XIS\,1 &   $0.954\pm0.005$\\
  & XIS\,3 &   $1.012\pm0.005$\\
  & HXD &   $1.215\pm0.009$\\
  & FPMA$\ast$ &  1\\
  & FPMB &  $1.030\pm0.004$\\
  CA$_\mathrm{Gal}$ & $N_\text{H,Gal}^\ast$~[$10^{22}\,\mathrm{cm}^{-2}$] &  0.023\\
  XSTAR 1 & $N_\text{H}$~[$10^{22}\,\mathrm{cm}^{-2}$] &  $1.2\pm 0.3$\\
  & $\log{\xi}~[\mathrm{erg}\,\mathrm{cm}\,\mathrm{s}^{-1}$] &  $2.82^{+0.10}_{-0.11}$\\
  CA$_1$ & $N_\text{H,int}$~[$10^{22}\,\mathrm{cm}^{-2}$] &  $21^{+4}_{-2}$\\
  cov. factor & f$_\text{cvr}$ &   $0.46^{+0.06}_{-0.05}$\\
  CA$_2$ & $N_\text{H,int}$~[$10^{22}\mathrm{cm}^{-2}$] &   $8.8^{+0.5}_{-0.6}$\\
  Abs. line & $E~[\mathrm{keV}]$ &   $8.17^{+0.11}_{-0.09}$\\
  & $\sigma~[\mathrm{keV}$] &   $0.34^{+0.14}_{-0.07}$\\
  & Depth~[$2\pi\,\sigma\,\tau_\mathrm{line}$] &   $0.0339^{+0.0124}_{-0.0020}$\\
  HXCOMP & norm &   $0.047\pm0.002$\\
  & $\Gamma^\ast$ &  1.72\\
  & k\,$T_\mathrm{e}^\ast~[\mathrm{keV}]$ &  399\\
  & k\,$T_\mathrm{bb}^\ast~[\mathrm{keV}]$ &  0.05\\
  Fe~K$\alpha$ & norm~[$\mathrm{Ph}\,\mathrm{s}^{-1}\,\mathrm{cm}^{-2}$] &   $\left(2.27^{+0.13}_{-0.14}\right)\times10^{-4}$\\
  & $E~[\mathrm{keV}]$ &  $6.394^{+0.005}_{-0.006}$\\
  Fe~K$\beta$ & norm~[$\mathrm{Ph}\,\mathrm{s}^{-1}\,\mathrm{cm}^{-2}]^\ast$ &  $0.12\times \mathrm{norm}_\mathrm{Fe\,K\alpha}$\\
  & $E^\ast~[\mathrm{keV}]$ &  7.1\\
  LP$_1$ & norm &   $6.9^{+1.2}_{-4.9}$\\
  & height~[$r_\mathrm{EH}$] &  $1.1000^{+0.0013}_{-0.0000}$\\
  & $i^\dagger$ &  $3^{+6}_{-0}$\\
  & $\Gamma^{\ast}$ &  1.72\\
  & $\log\xi$~[$\mathrm{erg}\,\mathrm{cm}\,\mathrm{s}^{-1}$] &   $2.835^{+0.016}_{-0.048}$\\
  & $Z_\mathrm{Fe}^\dagger$ & $2.5^{+0.6}_{-0.5}$\\
  & k$\,T_\mathrm{e}^\ast~[\mathrm{keV}]$ & 399\\
  LP$_2$ & norm &   $\left(8\pm 2\right)\times10^{-4}$\\
  & height~[$r_\mathrm{EH}$] &   $14.1^{+3.8}_{-1.9}$\\
  & $i^\dagger$ &  $3^{+6}_{-0}$\\
  & $\Gamma^\ast$ & 1.72\\
  & $\log\xi^\dagger$~[$\mathrm{erg}\,\mathrm{cm}\,\mathrm{s}^{-1}]$ &  2.84\\
  & $Z_\mathrm{Fe}^\dagger$ &  $2.5^{+0.6}_{-0.5}$\\
  & k$\,T_\mathrm{e}^\ast~[\mathrm{keV}]$ &  399\\
  SXCOMP & norm &  $\left(1.33\pm 0.05\right)\times10^{-3}$\\
  & $\Gamma^\ast$ &  1.72\\
  & k\,$T_\mathrm{e}^\ast~[\mathrm{keV}]$ &  399\\
  & k\,$T_\mathrm{bb}^\ast~[\mathrm{keV}]$ &  0.05\\
  \hline
\end{tabular}
\label{tab:bestfit}
\end{table}
\begin{figure*}
  \centering
  \includegraphics[width=0.48\textwidth]{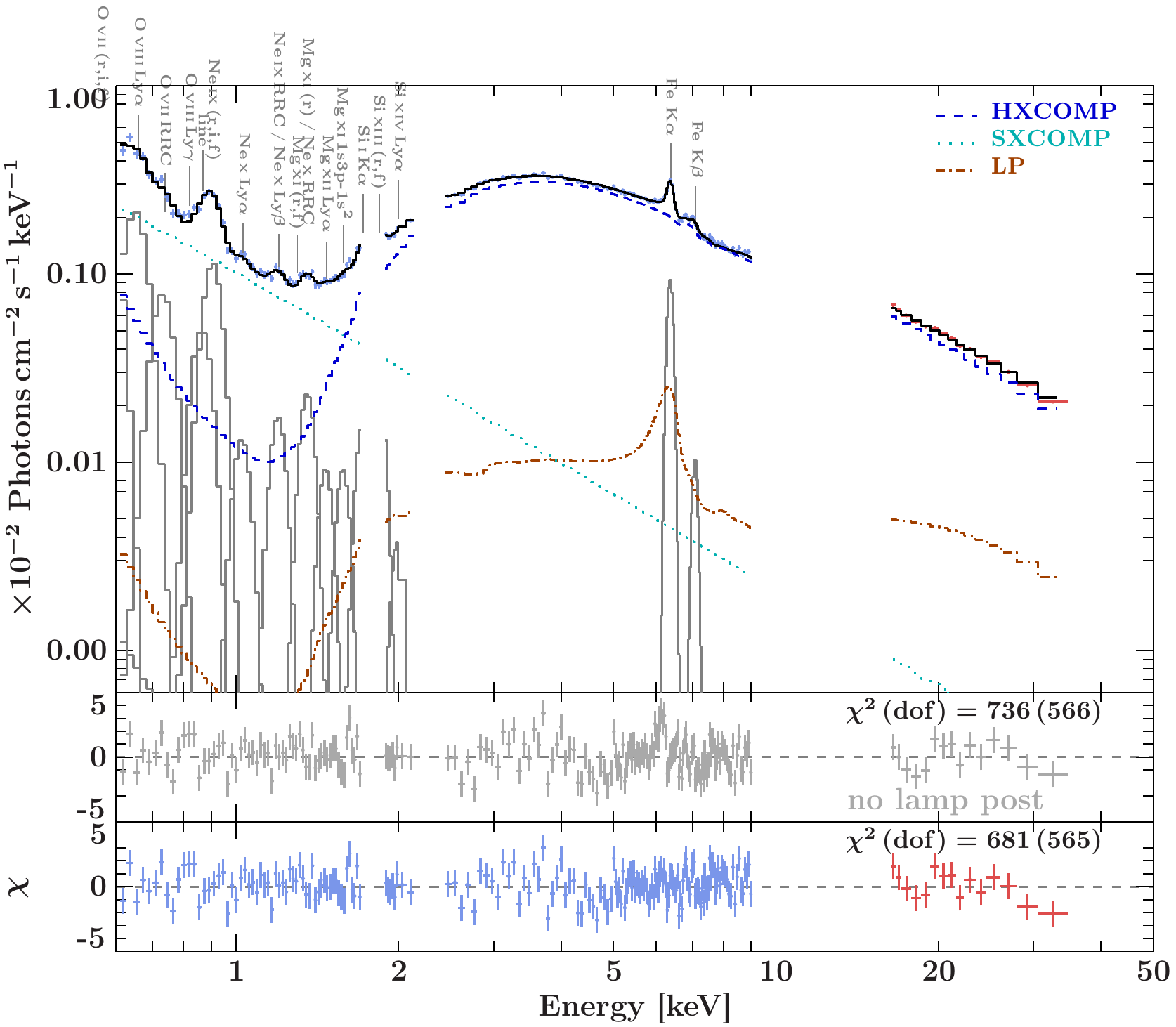}\hfill
  \includegraphics[width=0.48\textwidth]{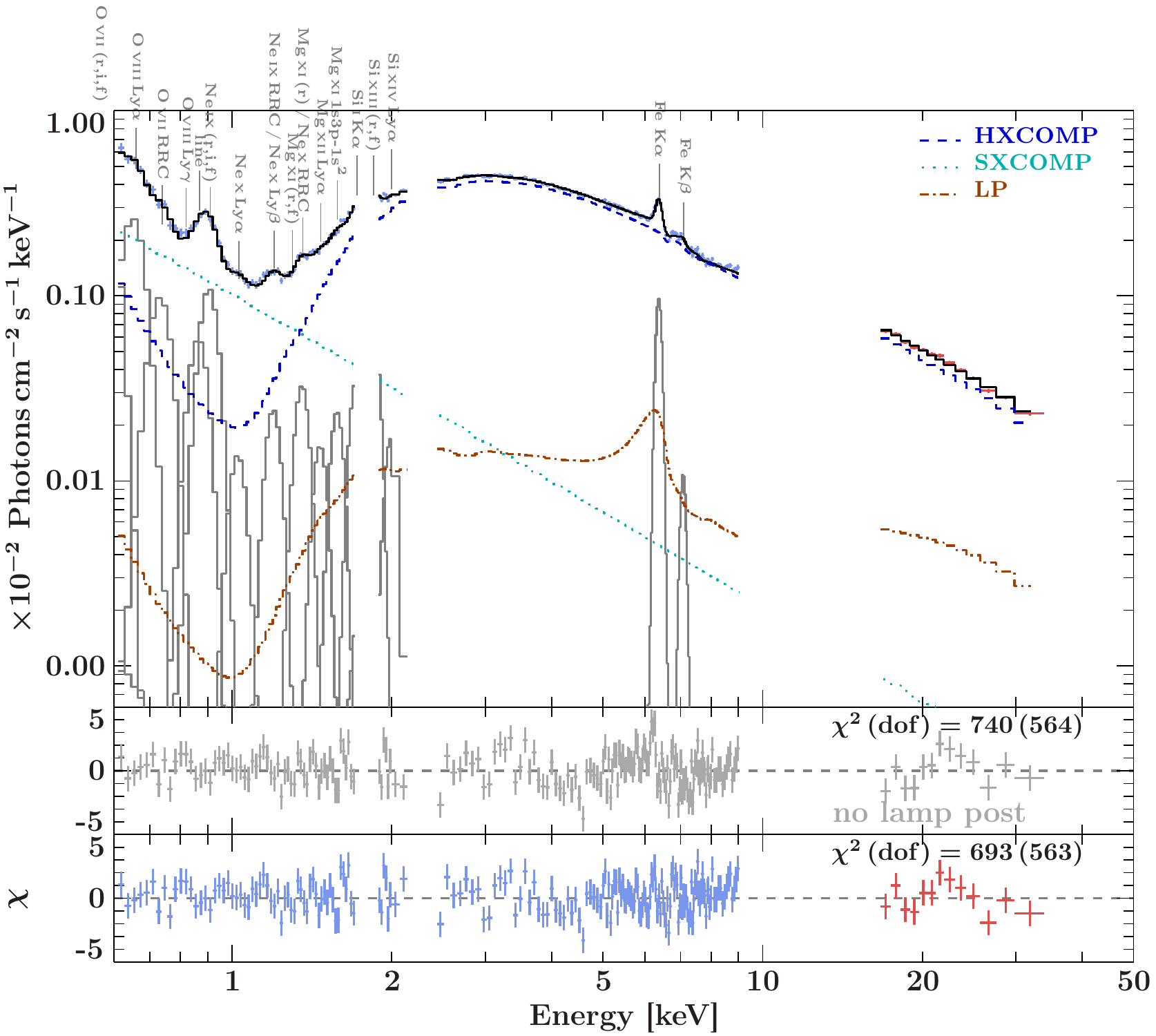}
  \caption{Model component plots for Suz~1 \textit{(left)} and Suz~2
    \textit{(right)} with the single lamp-post component drawn as
    brown dotted-dashed line. We also show the absorbed
    HXCOMP (blue dashed line) and the unabsorbed SXCOMP (cyan
    dotted line). The soft emission lines (gray) are labeled
    with their line identifiers. The residuals of \suzaku/XIS and
    \suzaku/HXD are drawn in blue and red, respectively. }
  \label{fig:modcomp_suz1_suz2_avrg}
\end{figure*}

We test for degeneracies inherent to the complex continuum for the
case of Suz~3 and study the resulting contours in $\Delta
\chi^{2}$-space (Fig.~\ref{fig:suz3_lp_contours}). The contours and
derived uncertainties suggest that we are able to constrain and
clearly separate between both lamp-post continua with primary sources
at different heights and a two-fold absorber, composed of a
partial-covering and full-covering neutral column.  We only observe a
tentative correlation between the LP$_2$-normalization and the column
density of CA$_1$. This correlation is likely not physical but caused
by both models describing a similar spectral shape around
6\,keV. We can show that the additional absorber of highly
  ionized gas is free of degeneracies with the continuum and can be
  kept separate throughout the analysis.

\subsubsection{The reflection fraction and reflection strength} 
\label{sec:reflfrac}
In our best-fit baseline model for Suz~3, we fit the primary continuum
independently from the reflection continua LP$_1$ and LP$_2$. The
normalizations of the reflection continua have no geometrical
interpretation in this case. In other words, this does not allow us to
infer the reflection fraction $R_\mathrm{f}$, which is intrinsic to
the lamp-post geometry and corresponds to the ratio of photons
escaping the system towards the observer over the coronal photons
incident on the disk \citep{Dauser2014,Dauser2016}. The reflection
fraction therefore mainly depends on the effect of light-bending,
which is stronger for primary lamp-post sources close to the disk,
that is, predicting larger values of reflection fraction $R_\mathrm{f}$
\citep{Miniutti2004b}. In the following, we use a workaround in order
to find access to the intrinsic reflection fraction for both LP$_1$
and LP$_2$. We re-define the baseline model, remove the
independent HXCOMP continuum and thaw $R_\mathrm{f}$ with
\texttt{refl\_frac$>0$}. This model describes the same spectral
components but now provides two primary continua, instead of one, that
are intrinsically linked to the reflected continua via the lamp-post
geometry, that is, it allows to directly fit for $R_\mathrm{f}$ as a more
meaningful parameter.

If we fit for both primary continua and both reflection fractions at a
time, we find strong degeneracies and large uncertainties with
$N_\mathrm{LP_1}=0.5^{+6.6}_{-0.4}$ and
$N_\mathrm{LP_2}=\left(1.68^{+0.11}_{-0.84}\right)\times10^{-3}$ as
well as $R_\mathrm{f}^\mathrm{LP_{1}}=14^{+36}_{-13}$ and
$R_\mathrm{f}^\mathrm{LP_{2}}=0.43^{+49.58}_{-0.05}$. For that reason,
we mutually freeze the reflection fraction of one component to the
value predicted for the emission of a point source at the previously
determined height above the disk (\texttt{fixReflFrac=2}) and fit for
the reflection fraction of the other component as well as both
normalizations of the primary continua. The results are shown in
Table~\ref{tab:reflfrac}. In the left column, LP$_2$ is fixed to a
predicted reflection fraction of 1.2 for the given height of
$h_\mathrm{LP_2}=14.1\,r_\mathrm{EH}\sim 15.0\,r_\mathrm{g}$.  A fit of
the reflection fraction of LP$_1$ is now well constrained as
$R_\mathrm{f}^\mathrm{LP_{1}}=1.50\pm0.06$, which is rather low with
respect to its low height of $h_\mathrm{LP_1}=1.1\,r_\mathrm{EH}$ and
the therefore predicted fraction of 22.5. This is absorbed by a much
larger normalization of LP$_1$ as opposed to LP$_2$, owing to strong
degeneracies between these parameters. In the second case, we set the
reflection fraction of LP$_1$ to $R_\mathrm{f}^\mathrm{LP_{1}}=22.5$
as predicted for its height
$h_\mathrm{LP_1}=1.1\,r_\mathrm{EH}=1.17\,r_\mathrm{g}$ and freely fit
$R_\mathrm{f}^\mathrm{LP_{2}}$. We find a reasonably low reflection
fraction of $R_\mathrm{f}^\mathrm{LP_{2}}=0.437\pm0.023$ as expected
for a primary lamp-post source at larger height. Also, the
LP-normalizations do not diverge as strongly as in the case before. We
can still demonstrate that degeneracies make it challenging to
interpret the reflection fraction as a probe of the lamp-post geometry
for the case of two interacting lamp-post sources.

While the reflection fraction can not simply be inferred from the
observed spectra without knowledge of the geometry
\citep{Dauser2014,Dauser2016}, the reflection strength $R_\mathrm{s}$
is defined as the strength of the Compton hump of the reflection model
with respect to the primary continuum, that is, the flux-ratio of the
reflected to the incident continuum in the 20--40\,keV energy band.
The derived numbers in Table~\ref{tab:reflfrac} imply a behavior very
similar to that observed for the reflection fraction. The reflected
LP$_1$-spectrum seems to be too weak with
$R_\mathrm{s}^\mathrm{LP_1}=0.69$ compared to the value of 10.5 as
predicted by the geometry of a point source very close to the black
hole, which is again accounted for by degeneracies between the
reflection fraction and the LP-normalizations. In contrast, we find a
reasonable value of $R_\mathrm{s}^\mathrm{LP_2}=0.24$ at a larger
height, featuring a rather weak Compton hump close to the predicted
value of 0.64, predicted for a point source at the given height. The
reflection strength of the combination of both lamp-post components is
0.68.

\begin{table}
\centering
\caption{Values obtained for reflection fraction $R_\mathrm{f}$, the
  reflection strength $R_\mathrm{s}$ and normalization of the two
  lamp-post components LP$_1$ and LP$_2$. Frozen
  parameters are denoted by an asterisk ($\ast$).}
\resizebox{0.9\columnwidth}{!}{
\begin{tabular}{ lll }
\hline\hline 
  &  free LP$_1$ & free LP$_2$ \\
  \hline
  norm LP$_1$ & $4.59\pm0.13$ & $0.306\pm0.011$ \\
  norm LP$_2$ & $\left(6.6\pm0.4\right)\times10^{-4}$ & $\left(1.822\pm0.019\right)\times10^{-3}$ \\
  $R_\mathrm{f}^\mathrm{LP_{1}}$ & $1.50\pm0.06$ & $22.5^\ast$  \\
  $R_\mathrm{f}^\mathrm{LP_{2}}$ & $1.2^\ast$ & $0.44\pm0.02$ \\
  $R_\mathrm{s}^\mathrm{LP_{1}}$ & $0.69$ & $10.5^\ast$  \\
  $R_\mathrm{s}^\mathrm{LP_{2}}$ & $0.64^\ast$ & $0.24$ \\ 
  \midrule
  $R_\mathrm{s}^\mathrm{LP_{2}+LP_{2}}$ & \multicolumn{2}{r}{$0.68$} \\
  \hline   
\end{tabular}
}
\label{tab:reflfrac}
\end{table}

\subsection{Test for relativistic reflection in other \suzaku\ and \xmm\ observations}
\label{subsec:relrefl_other_obs}
We  have now derived a robust baseline model based on the average
150\,ks observation by \suzaku\ (Suz~3) and \nustar\
(Nu$_\mathrm{Suz}$) that can also be applied to the remaining
observations of \suzaku\ and \xmm. We again use the original model
description with one primary continuum (HXCOMP) that is fitted
independently of the reflection continua (see
Table~\ref{tab:bestfit}). We re-fit the observations Suz~1 and
  Suz~2 and freeze all parameters except of the two absorbers CA$_1$
and CA$_2$ and the normalizations of the incident continuum and the
narrow Fe~K$\alpha$ and K$\beta$ lines. This approach yields overall
good fits, except for broad excess residuals in the iron band,
indicating extra variability of the reflection components. A
re-fit of the normalizations of the lamp-post continua results in
strong degeneracies. When fitting with both lamp-post components
switched off, we find the gray residuals shown in
Fig.~\ref{fig:modcomp_suz1_suz2_avrg}, arguing for the presence of
blurred reflection features via broad features in the iron band. We
can, however, demonstrate that the data of Suz~1 and Suz~2 do not
allow us to disentangle two emitting sources at different heights.
\begin{figure}
  \resizebox{\hsize}{!}{\includegraphics{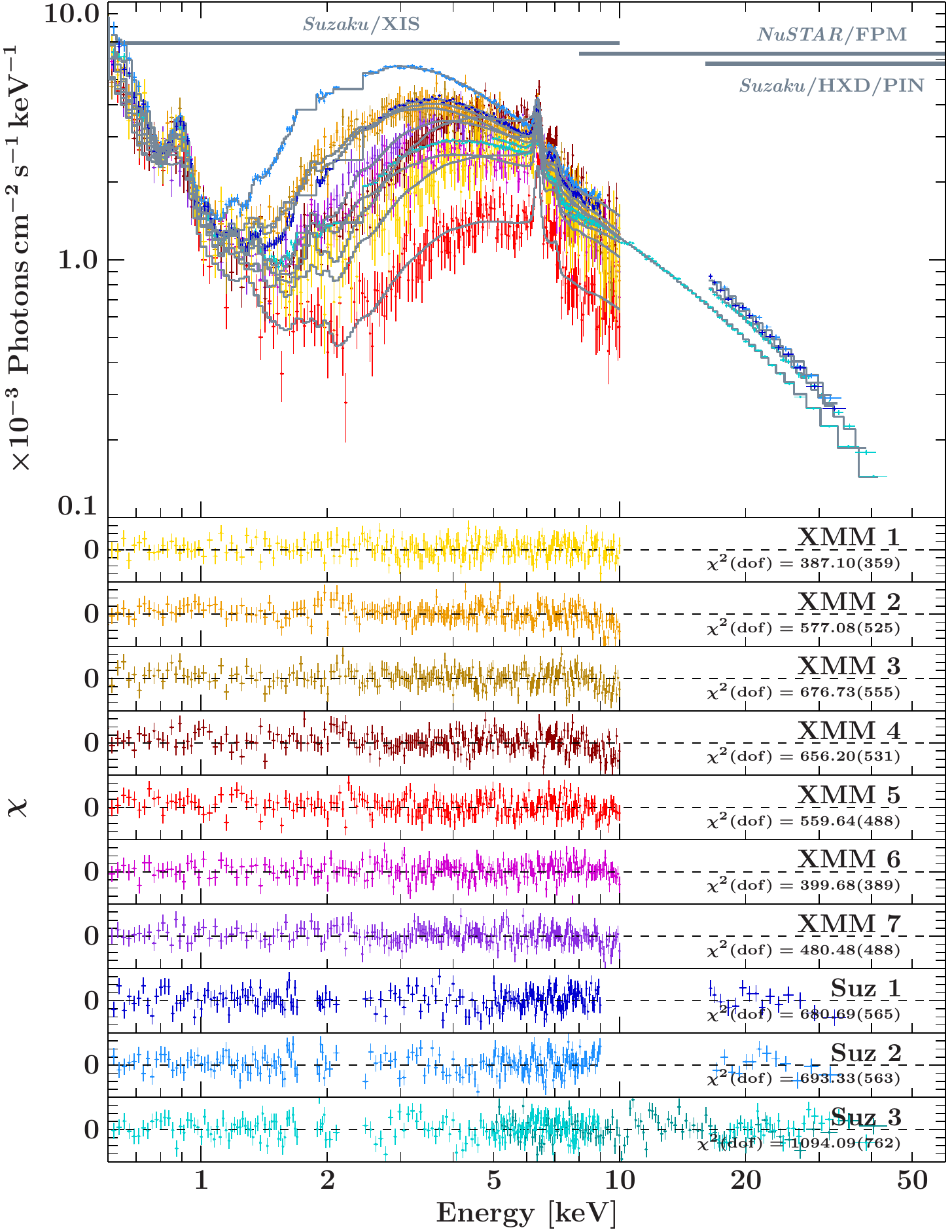}}
  \caption{Composite of all \xmm\ and \suzaku\ spectra as well as one
    \nustar\ spectrum simultaneous to Suz~3 with the applied best-fit
    baseline model and residuals. The statistics of each fit are
    listed in the residual panels. The minor ticmarks in the
      residual panels are separated by 0.1.}
  \label{fig:ngc4151_all}
\end{figure}
\begin{figure}
 \resizebox{\hsize}{!}{\includegraphics{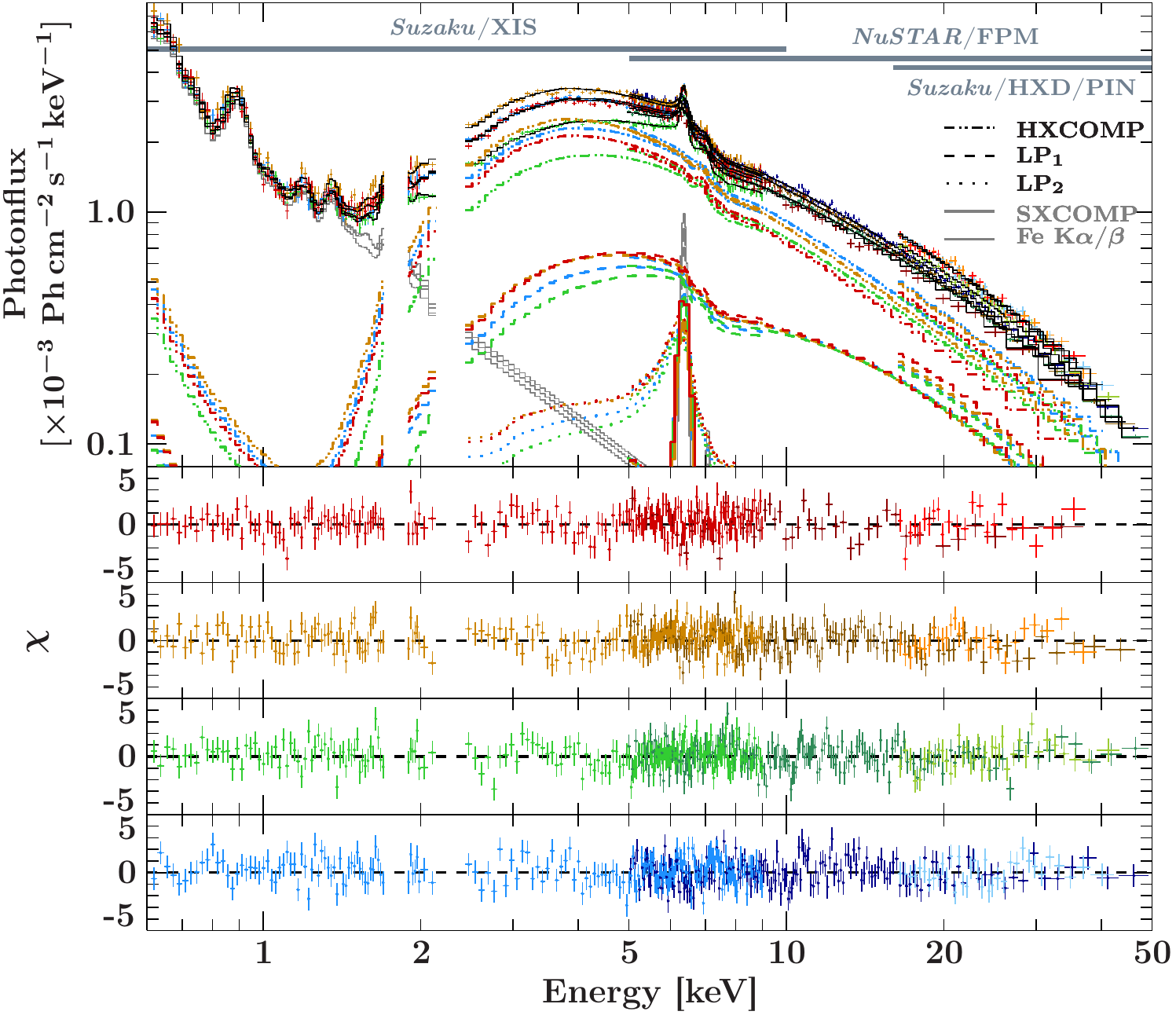}}
 \caption{Time-resolved spectra of
   Suz~3$_\mathrm{A}$--Suz~3$_\mathrm{D}$ /
   Nu$_\mathrm{Suz,A}$--Nu$_\mathrm{Suz,D}$ with the best-fit model
   components HXCOMP (double-dotted dashed line), LP$_1$ (dashed
   line), LP$_2$ (dotted line) and SXCOMP (solid line). The spectra
   A--D are drawn in red, orange, green and blue, respectively.}
  \label{fig:modcomp_suz3_sub}
\end{figure}

Given the comparatively large amount of counts for Suz~1 and Suz~2 we
attempt to constrain a single lamp-post component with variable
height. We find overall good fits
(Fig.~\ref{fig:modcomp_suz1_suz2_avrg}, bottom panels). The statistics
improve substantially compared to the model lacking relativistic
reflection with $\Delta \chi^{2}=55\,(1)$ for Suz~1 and with
$\Delta \chi^{2}=47\,(1)$ for Suz~2. The lamp-post heights are fitted
with $24^{+7}_{-9}\,r_\mathrm{EH}$ for Suz~1 and
$14^{+4}_{-6}\,r_\mathrm{EH}$ for Suz~2. Due to parameter degeneracies
we can neither state variability of this single LP-component between
Suz~1 and Suz~2, nor between the time-resolved spectra of Suz~2. A
direct comparison with the complex and well constrained
double-lamp-post source in Suz~3 are not possible either. We suggest,
however, that the single component fitted to Suz~1 and Suz~2
may be a blend of these components LP$_1$ and LP$_2$.

Similar to the \suzaku\ data, the \xmm\ data reveal visible broad
features in the iron-band residuals of a fit using the unblurred
\texttt{xillver} model (Fig.~\ref{fig:ngc4151_all_xillver_ironline}).
Due to the strong degeneracies arising for a free lamp-post height in
the models of Suz~1 and Suz~2, we freeze the heights of both lamp-post
components to those derived for Suz~3 and only fit the normalizations.
Table~\ref{tab:bestfit_var_xmm} shows that LP$_1$ is undetected by
\xmm\ in contrast to LP$_2$, where we can constrain its normalization
to within $\sim$12--60\% with the exception of a very good constraint
of $\sim $4\% for XMM~5. Together with the \xmm\ data, we report
significant variability in normalization for LP$_2$ over time with a
minimum timescale of 20--30\,d.

These results demonstrate the need for a good number of counts to
properly constrain one or even multiple components of relativistically
blurred reflection. Otherwise, it is challenging to simultaneously
probe the stability of both reflection components,
LP$_1$ and LP$_2$, over time.
\subsection{Spectral variability probed with \suzaku\ and \xmm}
\label{sec:specvar}
As we have shown in Sect.~\ref{sec:spectra}, NGC~4151 shows
significant spectral variability between 1 and 6\,keV, on which we will
focus in the following. We consider all \xmm, \suzaku, and
\nustar\ observations between 2011 and 2012 including the
time-resolved observations Suz~2$_\mathrm{A,B}$ as well as
Suz~3$_\mathrm{A-D}$ and Nu$_\mathrm{Suz,A-D}$.

To address the variability of the spectral components, we
apply the baseline model and allow only a few parameters to
vary. Besides the cross-normalizing detector constants and the flux
normalizations of the model components, these are \nha, \nhb, and the
covering fraction \fcov\ of CA~1.  We detect strong degeneracies
between \nha\ and \fcov\ for \xmm\ data and therefore fix the covering
fraction to the weighted mean with respect to the observations Suz~1,
Suz~2, and Suz~3, which all lie very closely to
$f_\mathrm{cvr}=0.46$. Note that the SXCOMP normalization is kept
frozen to the value derived for Suz~3 for all \suzaku\ observations
but allowed to vary for \xmm.  The derived values are, however,
consistent with the frozen value.

The baseline model fits well to all observations with only the few
above mentioned free parameters. All parameters and uncertainties are
listed in the Tables~\ref{tab:bestfit_var_xmm} and
\ref{tab:bestfit_var_suz}. Figure~\ref{fig:ngc4151_all} shows all
spectra with overlaid fits in the top panel and residuals in the
bottom panels.  Most previous studies have found the X-ray emission
below $\sim$1--2\,keV to be non-variable
\citep[e.g.,][]{Yang2001,deRosa2007}; \citet{Landt2015} report on
weakly-variable coronal \ion{O}{vii} emission.  In contrast,
\citet{Wang2010} find evidence for significant variability of the soft
continuum. Our multiple observations yield 0.6--1.0\,keV fluxes that
remain within $\sim$6\% of each\footnote{The amount of variability is
  calculated as the standard deviation over the average.} but underlie
large uncertainties.  This range contrasts with $\sim$20\%
  variability for the 7--10\,keV flux of the incident HXCOMP
  continuum. We find no direct correlation between the SXCOMP and
HXCOMP variability.  We can also exclude correlated variability
between the \ion{Ne}{ix} emission line and the incident 7--10 keV
flux.  Within the statistical uncertainties, the \ion{Ne}{ix} line
flux remains constant over the probed timescale.  Due to the CCD
resolution of our spectra and the blend of emission lines, it is not
clear if the weak flux variability within 0.6 and 1.0\,keV is due to
soft continuum, the \ion{Ne}{ix} flux, or both.  The narrow
Fe~K$\alpha$/$\beta$ line, in contrast, is stable over the monitored
time interval.
\begin{figure*}
\sidecaption
\includegraphics[width=12cm]{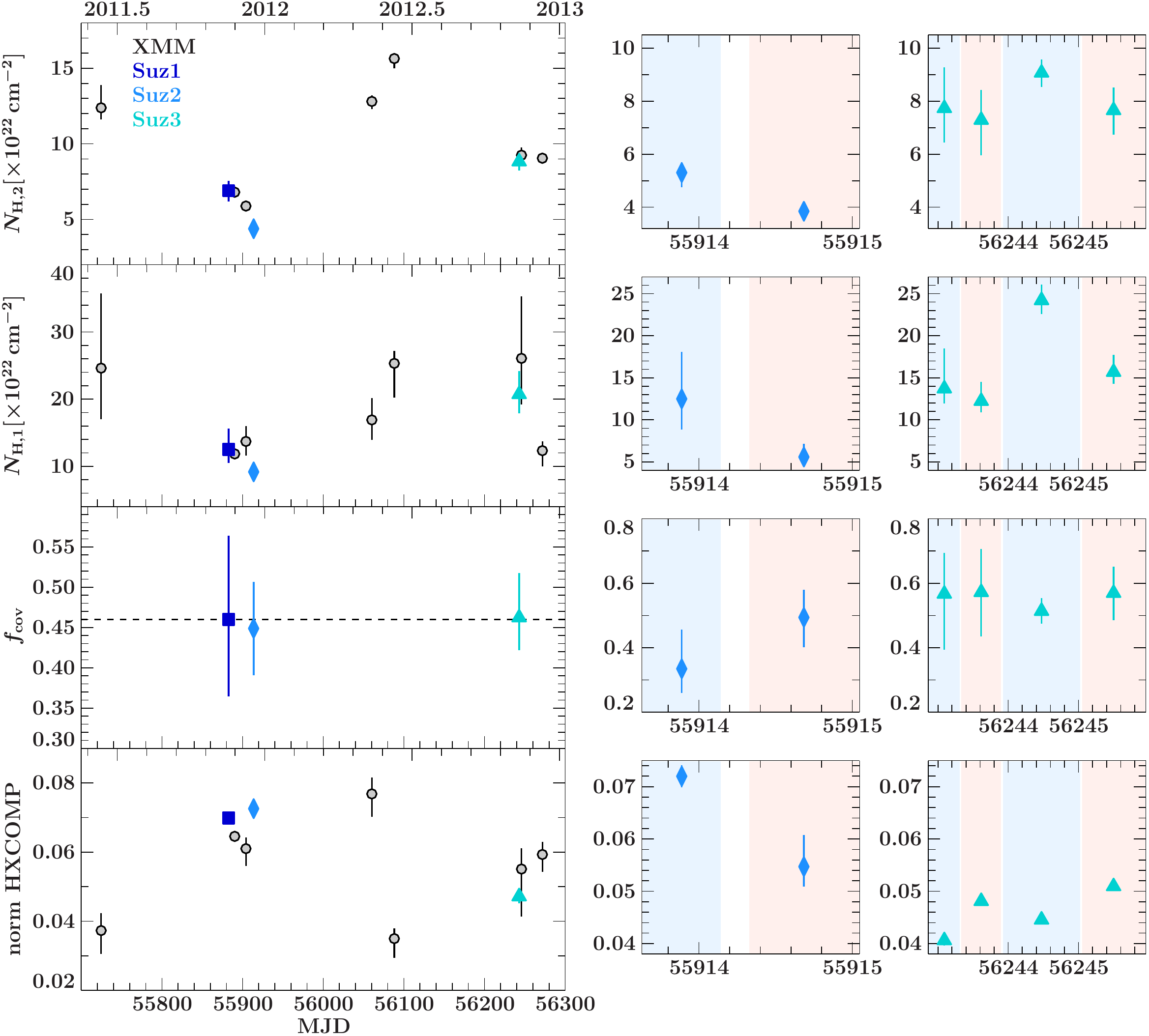}
\caption{\textit{Left:} Evolution of the free parameters related to
  the two absorbers and the normalization of the HXCOMP continuum.  We
  show parameters fitted to \xmm\ (gray) and the average \suzaku\
  spectra Suz~1, Suz~2, and Suz~3 (blue, light-blue, and
  turquoise). \textit{Right:} parameter evolution of the time-resolved
  spectra of the \suzaku\ observations Suz~2 (light-blue) and Suz~3
  (turquoise). The shaded regions correspond to those from
    Fig.~\ref{fig:gti_suz_hr}.}
\label{fig:varpars}
\end{figure*}

We find significant absorption variability with column density changes
of a maximum of 50\% for CA$_1$ and 20\% for CA$_2$.  Spectral
variability is not only found between single observations but also
within the 150\,ks-long observation Suz~3 as we show in
Fig.~\ref{fig:modcomp_suz3_sub}. The evolution of the parameters of
both neutral absorbers and the incident HXCOMP is shown in
Fig.~\ref{fig:varpars}. We find overall larger columns \nha\ but
smaller uncertainties for the partial coverer than for the
full-covering column \nhb. Both columns \nha\ (\nhb) seem to show
correlated variability over time with an initial decline around
MJD~55900 followed by an increase between MJD~56080 and MJD~56200 and
a subsequent decline after MJD~56240 back down to a baseline-level of
$N_\mathrm{H,1}\,(N_\mathrm{H,2})\sim 12\,(9)\times
10^{22}\,\mathrm{cm}^{-2}$. The normalizations of the primary
continuum (HXCOMP) can be shown to be variable down to a timescale of
$\sim$20\,d over a normalization range of $\sim
0.03$--$0.08\,\mathrm{Photons}\,\mathrm{cm}^{-2}\,\mathrm{s}^{-1}$ at
1\,keV.  The time-resolved measurements for Suz~2 and Suz~3 clearly
strengthen the presence of variability of these parameters on
timescales as short as days. This remains true even though the
uncertainties of the parameters found for Suz~2$_\mathrm{A,B}$ are
larger compared to those inferred from the spectra
Suz~3$_\mathrm{A-D}$. The covering fraction can be assumed as constant
within the uncertainties. The same applies to the time-resolved
results of Suz~3$_\mathrm{A-D}$, where we find covering fractions
scattering around 0.55. This value is slightly larger than that
derived for the total observation Suz~3, which can be attributed to
model degeneracies. The columns \nha\ and \nhb\ peak during the
observation Suz~3$_\mathrm{D}$ and cover a dynamic range of
$N_\mathrm{H,1}\,(N_\mathrm{H,2})\sim 13$--$25\,(6$--$10)\times
10^{22}\,\mathrm{cm}^{-2}$. The primary continuum, in contrast, shows
a more complex variability pattern with normalizations between
$0.04\,\mathrm{Ph}\,\mathrm{keV}^{-1}\,\mathrm{s}^{-1}\,\mathrm{cm}^{-2}$
and
$0.05\,\mathrm{Ph}\,\mathrm{keV}^{-1}\,\mathrm{s}^{-1}\,\mathrm{cm}^{-2}$.

\begin{figure}
  \resizebox{\hsize}{!}{\includegraphics{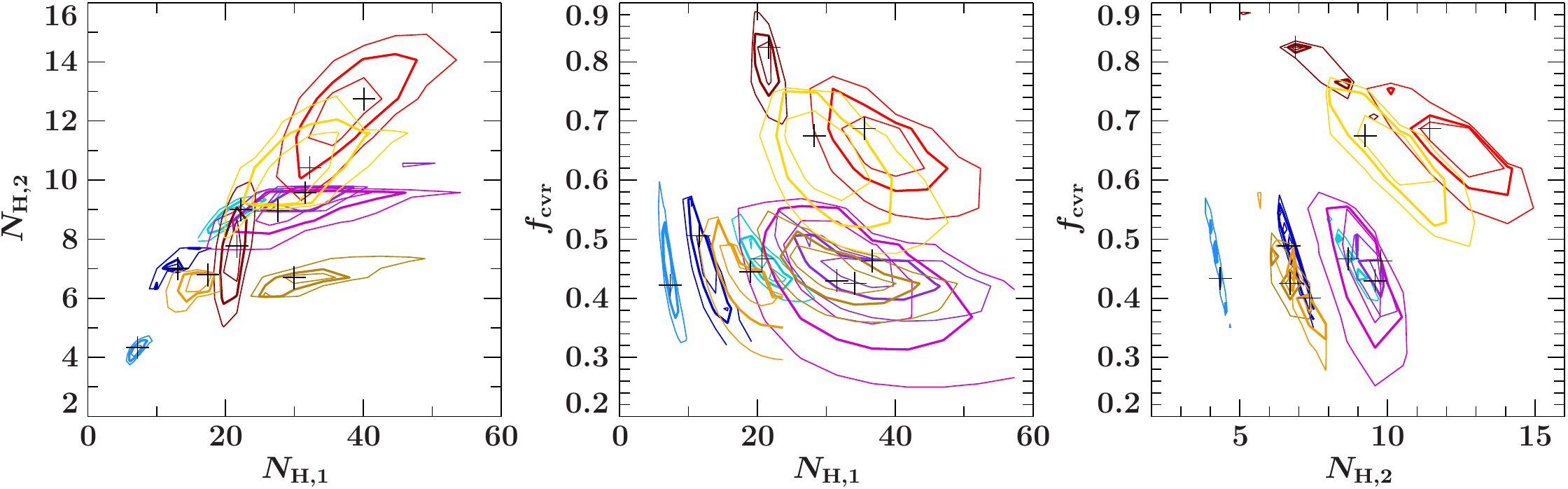}}
  \caption{Contours related to the two absorbers CA$_1$ and
    CA$_2$. The colors are the same used in
    Fig.~\ref{fig:ngc4151_all_xillver} and Fig.~\ref{fig:ngc4151_all}
    and correspond to the different observations. The best-fit
    parameters are indicated as black crosses.}
  \label{fig:contours}
\end{figure}

\begin{figure}
  \resizebox{\hsize}{!}{\includegraphics{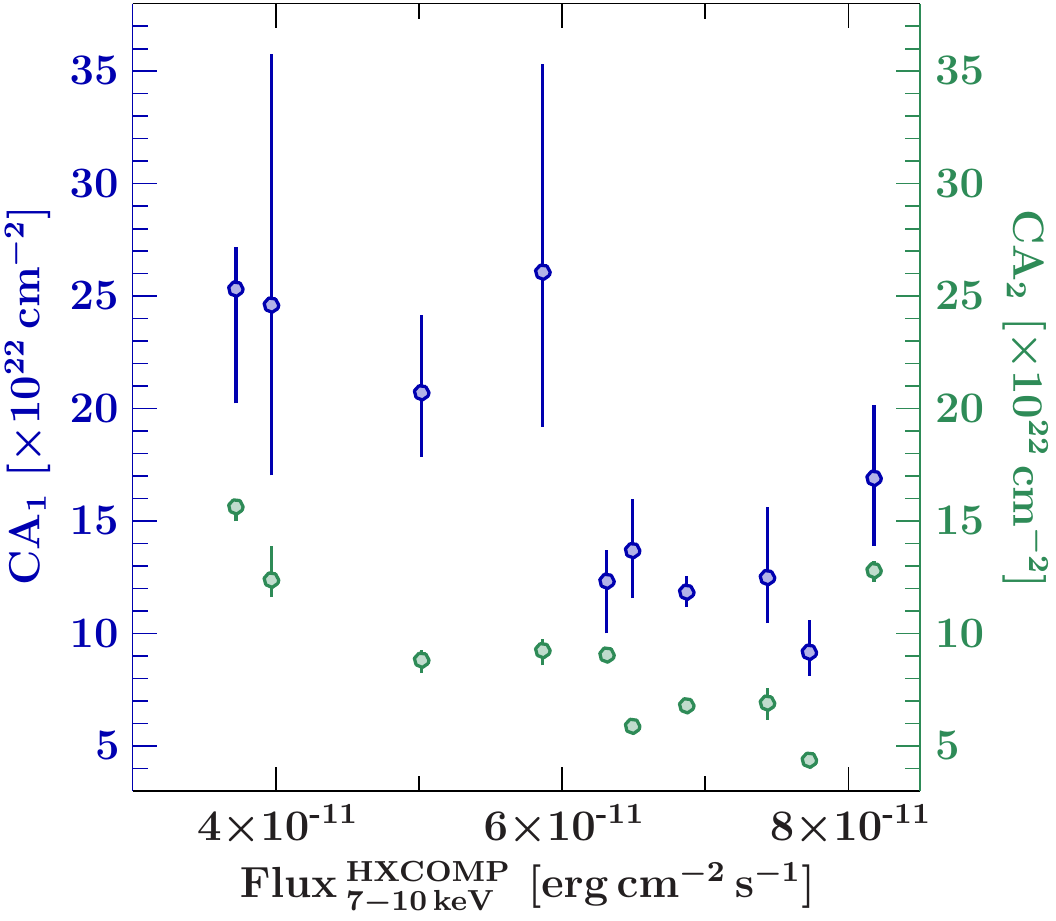}}
  \caption{Anticorrelation of the normalization of the irradiating
    flux (HXCOMP) with the column densities of the absorbers CA$_1$
    (blue) and CA$_2$ (green).}
  \label{fig:nh_hxcomp}
\end{figure}

The correlated variability between CA$_1$ and CA$_2$ on all probed
timescales can likely be attributed to degeneracies arising between
both absorbers (Fig.~\ref{fig:contours}) that reveal similar
columns. We probably also observe systematics related to the
complexity of the dual absorber and to the assumptions made for the
baseline model.  According to these results, we can therefore not
claim the observed variability to be inherent in one or the other
absorber for most of the analyzed observations. In contrast,
both contributing absorbers are likely well separated by the single
150\,ks-long observation Suz~3 (see Fig.~\ref{fig:suz3_lp_contours}).
We further observe an anticorrelation between the 7--10\,keV flux of
the pre-absorbed HXCOMP continuum and both columns \nha\ and \nhb,
which is highlighted in Fig.~\ref{fig:nh_hxcomp}. This relation is
stronger for CA$_1$ with the Pearson correlation coefficient
$r_\mathrm{P,CA_1}=77$\% and a very low p-value of
$P_\mathrm{P,CA_1}=0.8$\% as compared to CA$_2$ with
$r_\mathrm{P,CA_2}=60$\% and $P_\mathrm{P,CA_2}=7.2$\%.

\subsection{The soft X-rays and the NLR}
\label{subsec:softXrays}
\begin{figure*}
    \includegraphics[width=17cm]{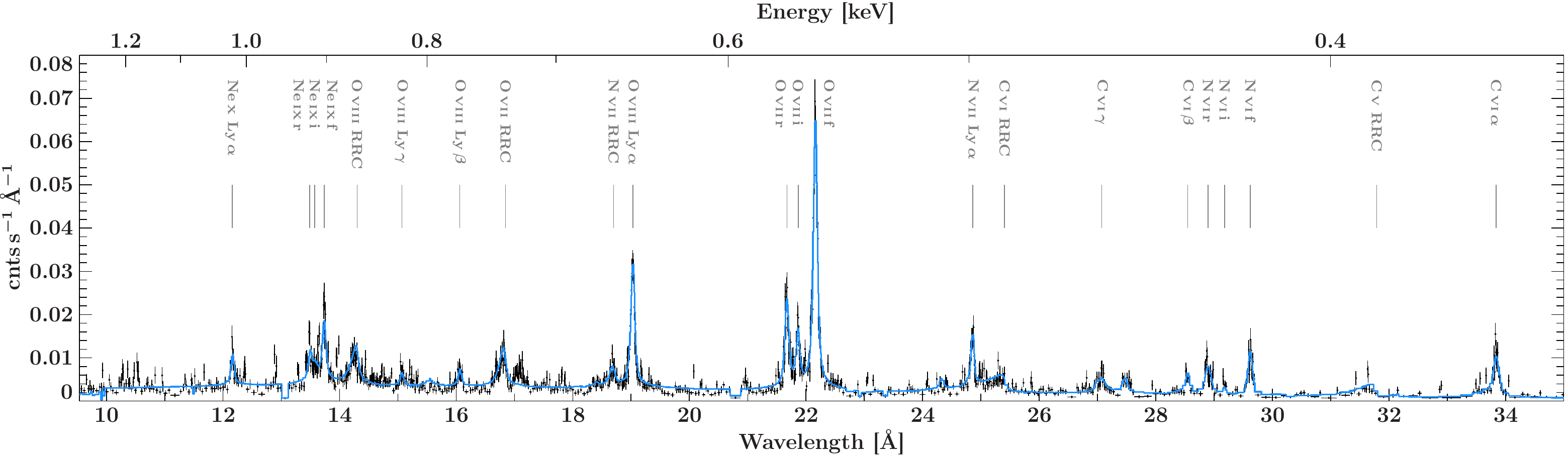}
    \caption{Combined count spectrum including all \xmm/RGS
      observations, each with two diffraction orders for each of the
      two RGS detectors RGS1 and RGS2. For reasons of visibility, the
      data are binned to a minimal S/N of 3. The model consists of the
      underlying SXCOMP continuum complemented with a set of emission
      lines in blue. Line identifiers are shown on top of the
      individual lines. }
  \label{fig:rgs}
\end{figure*}

NGC~4151 and its ionized environment have been subject to a number of
studies that combine high spectral and spatial resolution
\citep[e.g.,][and references therein]{Wang2011c}. In
Sect.~\ref{sec:suz3fit}, we motivate the description of the soft
X-rays below 2\,keV with the SXCOMP continuum, which is complemented
with a blend of Gaussian emission lines that have been measured to
persist over decades \citep[see also
][]{Ogle2000,Yang2001,Schurch2004,Wang2011c}. While
\citet{Perola1986}, \citet{Weaver1994a}, \citet{Weaver1994b},
\citet{Warwick1995} and \citet{Wang2010} detect no signs for major
variability in the soft X-rays, we find tentative signs for
variability at a low dynamic range in Sect.~\ref{sec:specvar}.  As the
S/N of individual RGS spectra are too low for measuring line fluxes
over time, we combine the \xmm/RGS data of all observations ranging
over more than 1.5\,years. The resulting spectrum reveals a series of
highly significant emission lines, dominated by the H-like and He-like
ions \ion{O}{viii}, \ion{O}{vii} , and \ion{Ne}{ix} (Fig.~\ref{fig:rgs})
on top of the weak SXCOMP continuum. The emission lines are modeled
with Voigt profiles. Whenever the S/N of a line-feature is too low to
constrain its centroid energy, we use the values from
\citet{Vainshtein1978} to set the starting parameters. Additional
radiative recombination continua (RRC) are described with the
\texttt{redge} model. The model is simultaneously fitted to all
observations with the resulting parameters listed in
Table~\ref{tab:lines_manualfits}. The centroid energies of all fitted
lines are consistent with those derived earlier, for example, by
\citet{Ogle2000} or \citet{Schurch2004}. The poor constraints on the
width and the turbulent velocities arise due to the choice of
the Voigt profile as the physically motivated description of the emission
lines, coming along with more free parameters than a Gaussian
approximation. Despite these increased uncertainties, the Voigt
profile shape is statistically required by the most prominent lines.
In contrast, we can roughly constrain the plasma temperatures of the
prominent RRC features.

\begin{figure}
    \resizebox{\hsize}{!}{\includegraphics{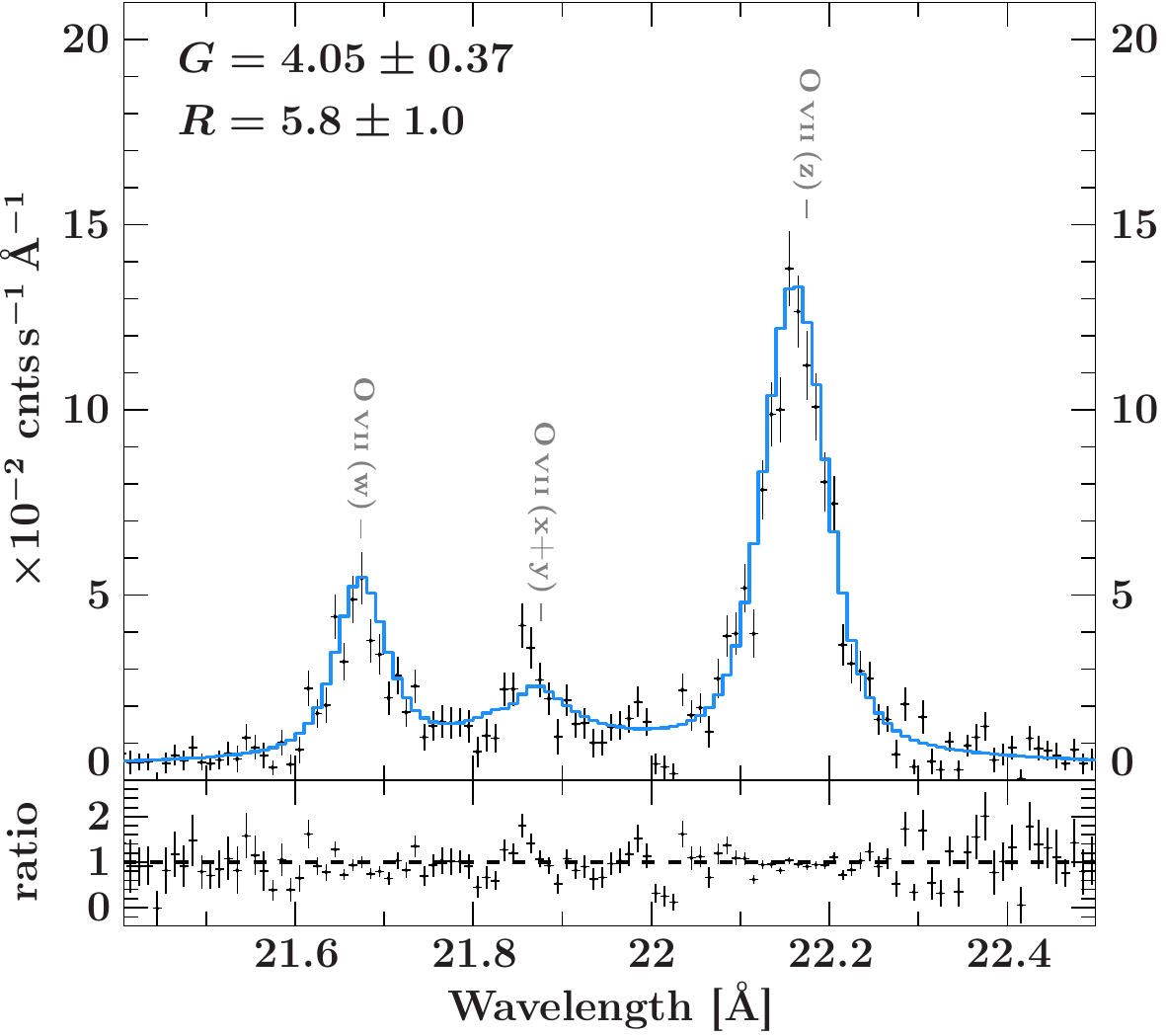}}
    \resizebox{\hsize}{!}{\includegraphics{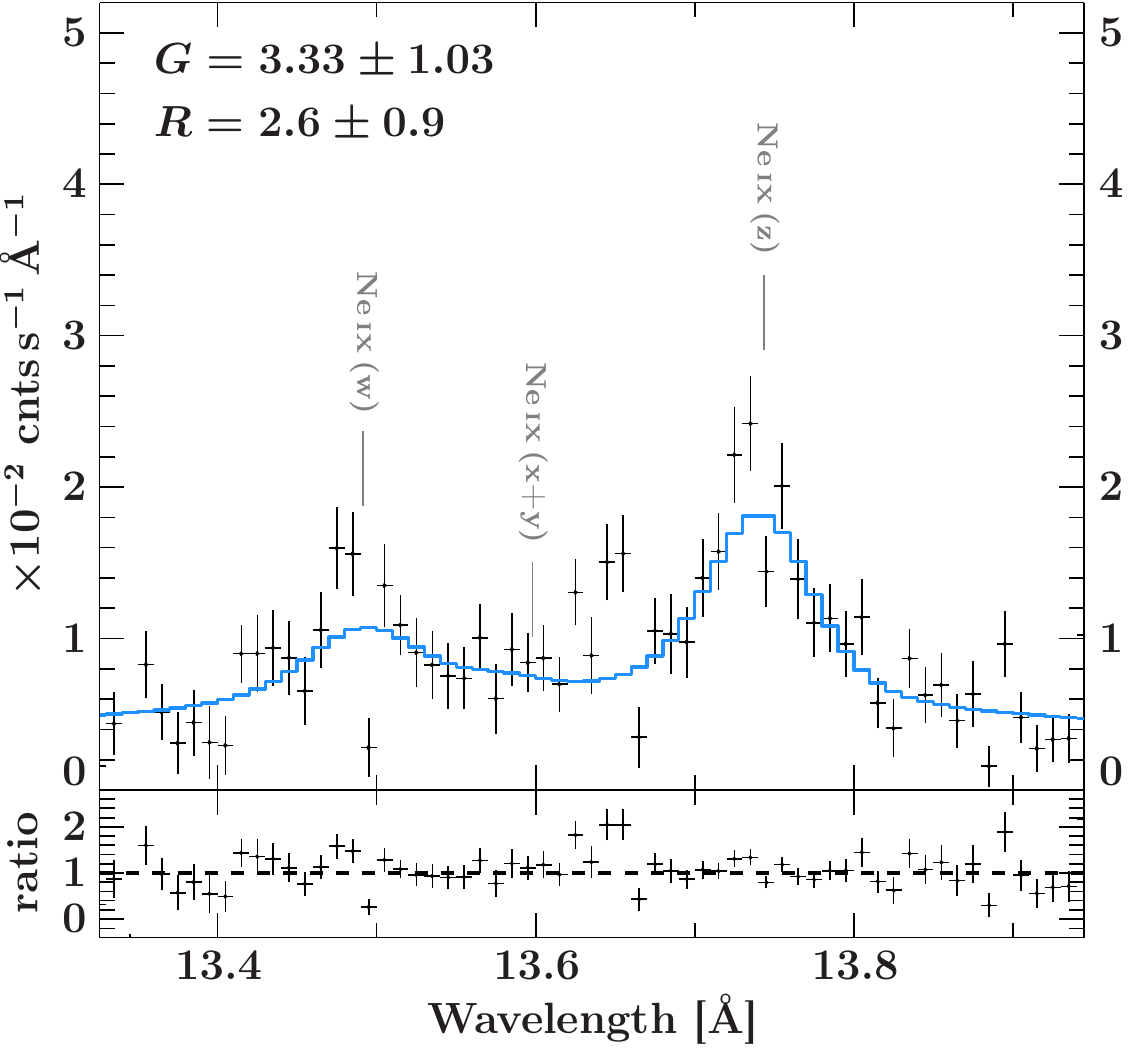}}    
    \caption{Enlarged view onto the bands of the He-like triplets of
      the ions \ion{O}{vii} (\textit{top}) and \ion{Ne}{ix}
      (\textit{bottom}). In both cases we show unbinned data from the
      first diffraction order combined out of all observations. The
      model corresponds to the parameters listed in
      Tables~\ref{tab:he_triplet} and \ref{tab:lines_manualfits}.}
  \label{fig:neix_var_timeres}
\end{figure}

A description of the soft X-rays using physical emission codes for
optically thin and ionized gas is beyond the scope of this paper,
whose main focus lies on a solid description of the nuclear
continuum. Also, the quality of the \xmm/RGS data does not allow for
improvement of results from previous studies with \chandra.  We therefore
remain with a phenomenological modeling of the soft X-rays and provide
a straight-forward gas diagnostic based on the two prominent line
triplets of \ion{O}{vii} and \ion{Ne}{ix} at $\sim$0.56\,keV (22\,\AA)
and $\sim$0.91\,keV (13.5\,\AA), respectively
(Fig.~\ref{fig:neix_var_timeres}). Both lines are statistically best
described by Voigt profiles.  We list the parameter constraints in
Table~\ref{tab:he_triplet}. In addition to the statistical
uncertainties, we also assume non-vanishing systematical uncertainties
for the fit of the hardly detected resonance and intercombination
lines of \ion{Ne}{ix}.

Both triplets consist of a resonance ($w$), an intercombination ($x+y$)
and a forbidden line ($z$), while the intercombination line is a blend
of the lines $x$ and $y$. Relating the strengths of these lines can be
used as a powerful diagnostic for the density and temperature of the
emitting gas \citep{Gabriel1969,Porquet2000,Bautista2000}. The ratio
$R=z/(x+y)$ is sensitive to the gas density in that the rate of
collisions increases with $n^2$, which suppresses the
forbidden line emission. The ratio $G=(z+x+y)/w$, in turn, is
sensitive to the temperature, which positions the gas between being
dominated by collisions or recombination. A hot plasma gives rise to
collisions and therefore a strong resonance line, that is, small values
of $G$. For larger values of $G\gtrsim 4$, the plasma is dominated by
recombination and the triplet levels with the intercombination and
forbidden line have large statistical weight, therefore featuring a
plasma dominated by photoionization. For both He-like ions, the ratios
are well consistent with those derived using \xmm/RGS by
\citet{Armentrout2007}. For \ion{O}{vii,} the ratios ($G=4.05\pm 0.37$
and $R= 5.8\pm 1.0$) indicate gas dominated by photoionization. These
ratios imply a low gas temperature of $T_\mathrm{e}\lesssim
10^{5}\,\mathrm{K}$, which is consistent with that obtained from the
narrow RRC features. In contrast to \citet{Schurch2004}, who find a
moderate $R$-ratio of 3.9, our value is more consistent with that
measured by \citet{Landt2015}, arguing for a gas of very low density
with $n_\mathrm{e}\sim 10^{3}\,\mathrm{cm}^{-3}$. We detect a
relatively strong resonance line for \ion{Ne}{ix}, resulting in a low
$G$-ratio of $3.333\pm 1.029$ and $R$-ratio of $2.6\pm 0.9$. This
result implies a hybrid plasma
\citep[e.g.,][]{Porquet2000,Bautista2000} that is likely in
pressure equilibrium with a collisional ionization gas phase
\citep{Wang2011c}. The temperature ($\sim 10^6\,K$) and density (a few
$\times 10^{11}\,\mathrm{cm}^{-3}$) are estimated to be significantly
higher compared to the \ion{O}{vii} gas. Again, the results are
overall consistent with earlier work
\citep{Schurch2004,Armentrout2007,Ogle2000,Wang2011c}. The suppression
of the forbidden line of \ion{Ne}{ix} and therefore the $R$-ratio can
alternatively arise from a strong UV field, where photoionization is
still dominant \citep{Mewe1978}.
\begin{table}
  \caption{Triplet line parameters of the ions \ion{O}{vii} and \ion{Ne}{ix}. Both lines are described with Voigt profiles. The thermal velocities are unconstrained and not listed in the table. The continuum parameters are frozen to those of the SXCOMP.}
  \resizebox{\columnwidth}{!}{
\begin{tabular}{llllll}
  \hline\hline 
  Line & E\,[keV] & FWHM\,[keV]  & flux\,[Ph\,s$^{-1}$\,cm$^{-2}$] \\
  \hline
  \ion{O}{vii}\,w & $0.57200^{+0.00013}_{-0.00006}$ & $\left(2.6^{+0.5}_{-1.1}\right)\times10^{-3}$ &  $\left(1.30^{+0.13}_{-0.10}\right)\times10^{-4}$\\
  \ion{O}{vii}\,x+y & $0.56702^{+0.00022}_{-0.00007}$ & $\left(2.7^{+1.7}_{-0.8}\right)\times10^{-3}$ & $\left(7.9^{+1.8}_{-0.7}\right)\times10^{-5}$\\
  \ion{O}{vii}\,z & $0.55946^{+0.00005}_{-0.00006}$ & $\left(2.5^{+0.6}_{-0.5}\right)\times10^{-3}$  & $\left(4.51^{+0.15}_{-0.13}\right)\times10^{-4}$\\
  \midrule
  \ion{Ne}{ix}\,w & $0.9191^{+0.0007}_{-0.0006}$ & $\le9\times10^{-3}$ & $\left(2.4^{+0.5}_{-0.9}\right)\times10^{-5}$ \\
  \ion{Ne}{ix}\,x+y & $0.9132^{+0.0013}_{-0.0015}$ & $\le 0.041$ & $\left(2.2^{+0.8}_{-0.7}\right)\times10^{-5}$\\
  \ion{Ne}{ix}\,z & $0.9024\pm0.0004$ & $\le6\times10^{-3}$ & $\left(5.8\pm0.6\right)\times10^{-5}$\\
  \hline
\end{tabular}
}
\label{tab:he_triplet}
\end{table}

\section{Discussion}
\label{sec:discussion}
Extending earlier work by \citet{Keck2015} by including soft X-ray
data, in this paper we re-investigated data from the joint
\suzaku/\nustar\ campaign of NGC~4151. The resulting broad energy
coverage and the use of the lamp-post model \texttt{relxillCP\_lp}
allowed us to disentangle complex and variable absorption from blurred
disk reflection originating in the regime of strong gravity. We
described the blurred reflection with two lamp-post components at
different heights, while the bulk of the 1--6\,keV continuum
absorption was attributed to an absorber consisting of partial- and
full-absorbing components. We also detected narrow absorption features
that can be modeled with a highly ionized ($\log\xi \sim 2.8$) warm
absorber and an additional absorption line indicating an UFO. The soft
emission was modeled with a SXCOMP scattering continuum and a number
of emission lines that are identified with \xmm/RGS.

\subsection{Size of the corona and pair production}
Our baseline model encompasses two point-like primary lamp-post
sources at the heights of $1.2\,r_\mathrm{g}$ and
$15.0\,r_\mathrm{g}$, which result in strongly and moderately blurred
spectral reflection components \citep[see also][]{Keck2015}.
Individual studies, for example, by \citet{Dauser2012}, \citet{Parker2014}
and Fink et al. 2017, in prep., that apply the model
\texttt{relxill\_lp}, confirm the need for relatively low source
heights between $\sim$2 and 4$r_\mathrm{g}$ and compact coronae in order
to explain the observed spectra.  This has been independently
confirmed by simulations of \citet{Svoboda2012} as well as
reverberation studies \citep[][for
  \mbox{1H\,0707$-$495}]{Fabian2009,Zoghbi2010}. In the following we
discuss the implications and limitations of this solution.

For the moment, we assume two distinct emission regions as they result
from our modeling. \citet{Dovciak2016} provide estimates on the
spatial extent of spherical coronae depending on their height above
the BH. For an observed luminosity of
$L_\mathrm{X,obs}^{0.3-10\,\mathrm{keV}}=0.001\,L_\mathrm{Edd}$ and a
photon index of $\Gamma=2$, the extent of a corona at a height of
10--$20\,r_\mathrm{g}$ grows to values larger than $1\,r_\mathrm{g}$.
The size is decreasing to $\sim 0.4\,r_\mathrm{g}$ at lower heights
and again increasing to $\sim 1\,r_\mathrm{g}$ for a height close to
$1\,r_\mathrm{g}$. This evolution of coronal sizes is caused by
light-bending close to the BH and gravitational redshifts at larger
heights. The models fitted to the \xmm, \suzaku,\ and \nustar\ data
examined in this paper imply the incident power law to have a mean
observed luminosity of
$L_\mathrm{X,obs}^{0.3-10\,\mathrm{keV}}/L_\mathrm{Edd}=0.001$--0.002,
which is comparable to the luminosity assumed by
\citet{Dovciak2016}. We therefore adopt a coronal radius of
$1\,r_\mathrm{g}$ for our components. If the corona of LP$_1$ was in fact situated at $1.2\,r_\mathrm{g}$ above the BH, its radius would
have to be even smaller in order not to interfere with the event
horizon.

The compactness of a corona with radius $d$ can be expressed as
\begin{equation}
  \ell=4\pi\,\frac{m_\mathrm{p}\,L_\mathrm{int}}{d\,m_\mathrm{e}\,L_\mathrm{Edd}}
  \label{eq:compactness}
,\end{equation}
where $L_\mathrm{int}$ is the intrinsically emitted luminosity,
$L_\mathrm{Edd}\sim 1.5\times
10^{38}\,(M/M_\mathrm{\astrosun})\,\mathrm{erg}\,\mathrm{s}^{-1}=4.5\times
10^{45}\,\mathrm{erg}\,\mathrm{s}^{-1}$ is the Eddington
luminosity\footnote{The Eddington luminosity is calculated with the
  black hole mass $M_\mathrm{BH}= 3\times
  10^{7}\,M_\mathrm{\astrosun}$ \citep{Hicks2008}.} and $d$
  is the radius of a simplified spherical corona in units of
  $r_\mathrm{g}$.  The compactness can be interpreted as the optical
depth of the corona with respect to pair production and has been
extensively discussed by \citet{Lightman1987}, \citet{Svensson1987},
\citet{Dove1997a} and \citet{Dove1997b}. A useful parameter in that
regard is $\Theta=k\,T_\mathrm{e}/m_\mathrm{e}\,c^{2}\sim
E_\mathrm{cut}/2\,m_\mathrm{e}\,c^{2}$ \citep{Garcia2015}, where
$E_\mathrm{cut}$ is the high-energy cutoff of the Comptonized
continuum. The cross-section for electron/positron pair production per
photon peaks between $E=m_\mathrm{e}\,c^{2}=511\,\mathrm{keV}$ and
$\sim$1\,MeV. We therefore evaluate the intrinsic luminosity for
the model of each observation both between 0.1 and 200\,keV (the energy
range used by \citealt{Fabian2015}) and 0.1 and 1000\,keV, which includes
the peak of the cross-section.

If we replace the model description of our best-fit baseline model
(Table~\ref{tab:bestfit}) with two separate incident continua and fit
for both normalizations ($N_\mathrm{LP_1}$, $N_\mathrm{LP_2}$) and
reflection fractions ($R_\mathrm{f}^\mathrm{LP_{1}}$,
$R_\mathrm{f}^\mathrm{LP_{2}}$), these four parameters are strongly
degenerate (see Sect.~\ref{sec:reflfrac} for the numbers). The dynamic
range of the normalizations comprises also the solutions of
Table~\ref{tab:reflfrac} and therefore provides the full range of
allowed values in a conservative way. The observed source luminosities
therefore range between $3.2\times
10^{41}\,\mathrm{erg}\,\mathrm{s}^{-1} \lesssim
L_\mathrm{obs,0.1-200\,keV}^\mathrm{LP_1} \lesssim 2.54\times
10^{43}\,\mathrm{erg}\,\mathrm{s}^{-1}$ and $1.17\times
10^{43}\,\mathrm{erg}\,\mathrm{s}^{-1} \lesssim
L_\mathrm{obs,0.1-200\,keV}^\mathrm{LP_2} \lesssim 2.50\times
10^{43}\,\mathrm{erg}\,\mathrm{s}^{-1}$. Especially in the case of
LP$_1$, where the corona would be situated deep inside the
gravitational potential, we need to transform the observed
luminosities into the intrinsic frame of the corona. The conversion
factor $(1+z_\mathrm{g})^\Gamma$ depends on the gravitational redshift
$z_\mathrm{g}=\left(1/\sqrt{1-2\,h/(h^{2}+a^{2})}\right)-1$ between the corona
and the observer as well as the photon index $\Gamma$. Fixing the spin
at $a=0.998$, we find that the conversion factor increases quickly
from $\sim$1.14 ($z_\mathrm{g}\sim 0.08$) at $h\sim
15.0\,r_\mathrm{g}$ to $\sim$37 ($z_\mathrm{g}\sim 7.2$) at $h\sim
1.2\,r_\mathrm{g}$. In addition, in the case of LP$_1$ we also need to
apply a correction factor to account for photon trapping. Due to
strong light-bending, at the height of $1.2\,r_\mathrm{g}$ only
$\sim$1\% of the photons of an isotropically emitting primary source
reach the observer, while a significant fraction of the photons
($\sim$13\%) reach the accretion disk, leading to a large
reflection fraction of $R_\mathrm{f}^\mathrm{LP_{1}}=22$ for LP$_1$
(see also Table~\ref{tab:reflfrac}). At this reflection fraction, the
corresponding luminosity becomes
$L_\mathrm{int,0.1-200\,keV}^\mathrm{LP_1}=0.4 L_\mathrm{Edd}$.
Leaving $R_\mathrm{f}$ free, the incident continuum can reach observed
luminosities as large as $2.54\times
10^{43}\,\mathrm{erg}\,\mathrm{s}^{-1}$. For this value, the corona
would intrinsically exceed $L_\mathrm{Edd}$ by a factor of ten.

The full range of allowed luminosities translates into a
compactness-range of $2500 \lesssim \ell_\mathrm{LP_1} \lesssim
200000$ and $5 \lesssim \ell_\mathrm{LP_2} \lesssim 11$. With these
constraints, we can show that the compactness parameter $\ell$ would,
within the uncertainties, well cover the parameter space below the
limits for pair production
\citep{Stern1995,Dove1997a,Dove1997b,Fabian2015} for LP$_2$. It would,
however, exceed this limit in some extreme cases of LP$_1$. Here, we
also have to correct the parameter $\Theta$ and the cutoff energy for
the gravitational energy shift arising between the compact source and
the observer. In the case of LP$_1$, this would shift the primary source
above the pair limit for all compactness values independent of the
value for $E_\mathrm{cut}$. As a result, the corona would be entirely
optically thick for photons with energies $E>511\,\mathrm{keV}$ with
respect to pair production.

Such extremely compact coronae have not been found in the sample of
AGN and XRBs studied by \citet{Fabian2015}. Also, the limitations that
we outline above and that are also mentioned by
\citet{Niedzwiecki2016} challenge the interpretation of LP$_1$ as
a distinct component. The component LP$_1$ can therefore only be
interpreted in combination with LP$_2$. Although the superposition of
the two distinct lamp-post components LP$_1$ and LP$_2$ serves well to
describe our data, we hesitate to claim the two point sources to
represent a realistic description of the corona. Also, the
strong degeneracies between the largely uncertain normalizations of
both lamp-post components and the low reflection fraction
found for LP$_1$ argue against two extremely compact and
distinct coronae.  In contrast, a continuous structure enclosing both
components seems more likely. The detected spectral signatures of
strongly blurred reflection yet indicate in a model-independent way
that at least part of the corona must lie very close to the black
hole. The low reflection fraction measured for LP$_1$ also
points towards a vertically extended structure \citep{Keck2015}. It
may be reflected by outflows \citep[see also][]{King2017} that have
been detected for NGC~4151
\citep{Kraemer2005,Tombesi2010,Tombesi2011}, where relativistic
aberration \citep{Beloborodov1999,Malzac2001} can effectively reduce
the observed fraction of reflected photons. These outflows together
with its non-relativistic jets
\citep[e.g.,][]{Pedlar1993,Ulvestad2005} could well fit in this
picture and may provide a natural environment for the source of
primary X-ray photons
\citep{Markoff2004,Markoff2005,Wilkins2015,King2017}.

Reverberation studies have independently suggested horizontally
\citep{Wilkins2012} and, in particular, vertically extended primary
sources above the black hole for Ark~564 \citep{Zoghbi2010}, Mrk~335
\citep{Kara2013}, IRAS~13224$-$3809 \citep{Wilkins2015} and also for
NGC~4151 \citep{Zoghbi2012,Cackett2014}, which indicates a primary
source within $\sim$5--$10\,r_\mathrm{g}$ from the black hole. This
may, in conjunction with the high-quality spectral information of the
long-look \suzaku\ and \nustar\ data, be well in agreement with a
jet-based geometry, which can still be relatively radially compact. Magnetic fields \citep[e.g.,][for the extreme case of the
radio galaxy NGC~1052]{Baczko2016} make this region an efficient
emitter of synchrotron photons
\citep[e.g.,][]{Merloni2000,Markoff2005}, which can act as additional
seed photons for Comptonization processes in the corona and therefore
at least reduce its transverse extent \citep{Dovciak2016} close to or
below the value of $1\,r_\mathrm{g}$, which we have previously assumed
for our estimates of the compactness.

\subsection{Complex absorption variability}
We model four layers of absorption: two neutral layers, one partially
covering the nucleus with $f_\mathrm{cvr}\sim 0.5$ (CA$_1$/\nha) and
one full-covering absorber (CA$_2$/\nhb), a warm absorber with $\log
\xi \sim 2.8$ as well as an additional broad absorption line around
8\,keV that can either be formed by a \ion{Fe}{xxvi}~Ly\,$\beta$ line
that is unmodeled by the warm absorber or, more likely, a blueshifted
\ion{Fe}{xxv}~He\,$\alpha$ or \ion{Fe}{xxvi}~Ly\,$\alpha$ line,
indicative of an UFO at a speed of $\sim$0.16--0.24\,c.

We showed that one or both of the absorbers CA$_1$ or CA$_2$ account
for the bulk of the spectral variability between 1--6\,keV on
timescales from days to years. This agrees with \citet{Puccetti2007}
and \citet{deRosa2007} for a very similar set of absorbers. The
partial-covering column \nha\ ranges between $10$  and $25\times
10^{-22}\,\mathrm{cm}^{-2}$. Its covering fraction averages 46\%
across the \suzaku\ observations. The full-covering column \nhb, on
the other hand, ranges between $5$ and $15\times
10^{-22}\,\mathrm{cm}^{-2}$. The columns \nha\ and \nhb\ seem to track
one another well; given the data quality and the comparable values of
both columns, we can explain this model-dependent result as being due to
degeneracies between \nha\ and \nhb. In agreement with earlier
analyses using \textit{BeppoSAX} \citep{Puccetti2007,deRosa2007} and
\chandra \citep{Wang2010}, we can therefore not tell if one or the
other column dominates the variability. A separation of both neutral
absorbers from the inner-disk reflection component is equally
challenging, as both describe very similar spectral features at the
flat turnover. We could still show that disentangling both
is possible with the high count statistics and broad spectral
coverage of the long-look \suzaku\ and \nustar\ observations as well
as the variable nature of the absorbers. The latter has also been
demonstrated by \citet{Risaliti2009b} and \citet{Risaliti2013} for
NGC~1365, which is very similar to NGC~4151 in that regard.

The shortest variability timescale of 2\,d has been measured with the
time-resolved analysis of Suz~3 (see Fig.~\ref{fig:varpars}). The
circumnuclear gas in NGC~4151 may be continuous yet non-homogeneous
and/or containing (or consisting of) a discrete number of localized
clouds. Such scenarios have been explored for this source in the past
by \citet{Holt1980}, \citet{Yaqoob1991}, \citet{Zdziarski2002}, and
\citet{Puccetti2007}.

In the case where the {variable} absorber consists solely of
clouds, variability either by $N_\mathrm{H}$ or the covering fraction
may be associated with single clouds entering or exiting the
line-of-sight. We can not exclude variability of the covering fraction
due to inherent degeneracies that lead us to fix this parameter. The
model by \citet{Nenkova2008b} attempts to explain line-of-sight
variability for all inclinations with a Poissonian distribution of
clouds that decrease in number density further out. This model has
been successfully fitted to infrared (IR) SEDs by \citet{AlonsoHerrero2011} and
is able to explain absorption events observed in the X-rays
\citep[][and, e.g., \citealt{Beuchert2015}]{Markowitz2014}. On
average, these studies predict only a few clouds on the line-of-sight
for inclinations similar to the the one we measure for NGC~4151. This
is consistent with independent estimates for NGC~4151 by
\citet{Holt1980}, \citet{Yaqoob1991} and \citet{Zdziarski2002}.

Regardless of whether the variable absorber is CA$_1$ or CA$_2$, we
can use the observed variability timescales of \nh\ to estimate the
location of potential clouds that are moving on Keplerian orbits
\citep{Risaliti2002,Puccetti2007} with
\begin{equation}
  \label{eq:absdist}
  R= 3.6\times 10^{17}\,\frac{M_\mathrm{BH}}{10^{7}\,M_\mathrm{\astrosun}}\left(\frac{n_\mathrm{H}}{10^{9}\,\mathrm{cm}^{-3}}\right)^{2}\left(\frac{\Delta t}{2\,\mathrm{d}}\right)^{2}\left(\frac{N_\mathrm{H}}{10^{22}\,\mathrm{cm}^{-2}}\right)^{-2}\,\mathrm{cm}
,\end{equation}
where $M_\mathrm{BH}$ is the black hole mass and where $n_\mathrm{H}$
and $N_\mathrm{H}$ are the cloud number density and column density,
respectively. For $\Delta t$ we use the shortest variability
timescale of 2\,d. The column density is set to the maximum modeled
value of $N_\mathrm{H,1}=\sim 25 \times 10^{22}\,\mathrm{cm}^{-2}$.
The number density is unknown. If we assume number densities for BLR
clouds of $n_\mathrm{H}=10^{9-10}\,\mathrm{cm}^{-3}$
\citep{Netzer1990,Kaspi1999,Netzer2008}, we obtain a distance range of
$R\sim 5.6\times 10^{-4}$--$5.6\times 10^{-2}\,\mathrm{pc}$ or
$3.9\times 10^{2}$--$3.9\times 10^{4}\,r_\mathrm{g}$. These distances
are consistent with those inferred by \citet{Puccetti2007} and with
the distance of the BLR \citep[$\sim 8\times
10^{-3}\,\mathrm{pc}$,][]{Maoz1991}. Note that for the inclination of
NGC~4151, the average number of clouds on our line-of-sight stays
approximately the same even if we extrapolate the clumpy torus model
down to the BLR. In contradiction to this theoretical consideration,
\citet{Arav1998} find no signs for distinct BLR clouds in
high-resolution optical Keck spectra. This result favors an
interpretation of our data with irregular and dynamic absorbing
structures. In theory, dust must be entirely or at least partially
sublimated at the inferred distances with the dust sublimation radius
\begin{equation}
  \label{eq:dustsubl}
  R_\mathrm{d}=0.13\,\mathrm{pc}\,
  \left(\frac{L_\mathrm{bol}}{10^{44}\,\mathrm{erg}\,\mathrm{s}^{-1}}\right)^{0.5}\left(\frac{T_\mathrm{d}}{1500\,\mathrm{K}}\right)^{-2.6}
,\end{equation} 
\citep{Nenkova2008b}. We find $R_\mathrm{d}\sim
0.13\,\mathrm{pc}=8.9\times 10^{4}\,r_\mathrm{g}$ with the assumed
bolometric luminosity
$L_\mathrm{bol}=10^{44}\,\mathrm{erg}\,\mathrm{s}^{-1}$
\citep{Vasudevan2009} and the dust evaporation temperature of
$T_\mathrm{d}=1500\,\mathrm{K}$ \citep{Barvainis1987}. When the inner
range of the torus is estimated independently, slightly smaller
distances are found; using the 5100\AA\ line luminosity
\citep{Kaspi2004}, the outer BLR of NGC~4151 could be constrained to a
distance of $\sim 6\times 10^{3}$--$6\times
10^{4}\,r_\mathrm{g}$, and the inner rim of the torus to $\sim$0.04\,pc
using thermal dust reverberation studies by \citet{Minezaki2004} and
\citet{Burtscher2009}. \citet{Schnuelle2015}, however, find no signs
for dust sublimation in their data and explain this with large
graphite dust grains that sublimate at much higher temperatures. If we
assumed clouds at a distance of $R=R_\mathrm{d}$, the cloud density
required by the observed $N_\mathrm{H}$ variability pattern would be
on the order of $\sim 6\times 10^{10}\,\mathrm{cm}^{-3}$, which can be
excluded for a dusty torus \citep{Elitzur2007}.

On the other hand, we also detect variability over longer timescales of
approximately one year that may indicate clouds at larger distances.
If we assume, for example, number densities of
$10^{7-8}\,\mathrm{cm}^{-3}$, which are typical for the dusty torus
\citep{Miniutti2014,Markowitz2014}, we find $R\sim 2\times
10^{-3}$--$2\times 10^{-1}\,\mathrm{pc}$ or $1.3\times
10^{3}$--$1.3\times 10^{5}\,r_\mathrm{g}$. These values
put the clouds into the outer BLR or at the inner side of the dusty
torus, which is also a common result of \citet{Markowitz2014} for
similar timescales. Much larger distances of clouds from well inside
the torus would require unrealistically large number densities and are
therefore unlikely.

Recently, \citet{Couto2016} published a study that investigates X-ray
absorbers with archival long-look \chandra\ observations. They find a
highly ionized column similar to our \texttt{xstar} component as well
as an outflowing near-neutral absorber with a speed of
$\sim500\,\mathrm{km}\,\mathrm{s}^{-1}$. This neutral absorber goes
back to absorption line features in \hst/GHRS/STIS data first
mentioned by \citet{Weymann1997} and referred to as the kinematic
component ``D+Ea'' in Table~1 of \citet{Kraemer2001}; see also
\citet{Kraemer2006} for further usage. \citeauthor{Couto2016} fit
these absorbers to a number of seven archival \chandra\ observations
with two additional observations by \xmm\ in 2000 and \suzaku\ in
2006. They conclude that the bulk of the spectral variability over
14\,years is caused by a change in the ionization state of both
absorbers as a response to changes in the irradiating luminosity
rather than the observed variations in \nh. Similar to the absorbers
used by \citet{Couto2016}, we find the columns of both neutral
absorbers, CA$_1$ and CA$_2$, to be anticorrelated with the incident
photon flux of the HXCOMP component. A portion of either of these is
likely consistent with the component D+Ea. Its outward motion forms a
consistent picture with the observed anticorrelation of the incident
flux with the column density, in which a still dusty, radiatively
driven wind \citep[e.g., ][]{Czerny2011,Dorodnitsyn2012} causes a
decrease of the line-of-sight absorption for stronger radiative
driving. We, however, emphasize that the interpretation of the data
analyzed in this work strongly depends on the method of modeling. Both
the interpretation with orbiting clouds or with an outflowing
near-neutral absorber represent structural changes in the absorber. We
do not favor one over the other. In a yet different scenario, the
ionization state of the absorber can change with varying irradiation,
resulting in changes of the equivalent column density. We, however,
are not sensitive to this effect with the available count statistics.

On larger scales, \citet{Ruiz2003} and \citet{Radomski2003} detect
dusty extended gas in NIR/MIR data, which is fully covering the
nucleus similar to dust that has been found within $\sim$4\,pc using
the Gemini NIR integral field spectrograph \citet{Riffel2009}. This
dusty component may therefore make up a non-variable portion of our
CA.

\subsection{The soft X-rays}
We model the soft X-rays of all observations with a SXCOMP with the
photon index tied to that of the HXCOMP as well as a blend of Gaussian
lines that are motivated from high-resolution grating observations of
\xmm/RGS and \chandra/HETG. We have shown that the soft flux below
1\,keV is, contrary to the hard flux, only mildly variable at a low
dynamic range of $\sim$6\% at most. The origin of the
  variability (continuum, emission line(s), or both) is unclear. If
we calculate the ratio of the SXCOMP and HXCOMP normalization, we can,
on average, infer a low optical depth of the soft emitting gas of
$\tau\sim 0.023$.

Due to the comparatively large PSFs of \xmm\ and \suzaku, we are
unable to spatially resolve the line-emitting gas. In contrast, a
number of authors have been using \chandra\ for this purpose, which is
both powerful in spatial and spectral resolution. \citet{Ogle2000} and
\citet{Wang2011c} show that a considerable part of the soft X-rays is
due to distant, extended gas that is spatially coincident with a
bi-conical gas distribution \citep[see][for integral field
spectroscopy of \ion{[O}{iii]}]{StorchiBergmann2010} and the
narrow line region (NLR, \citealt{Bianchi2006}). A spatially
resolved modeling of the extended gas with \texttt{Cloudy} allows
\citet{Wang2011c} to conclude a two-phase photoionized medium of
intermediate ($\log \xi\sim 1.7$) and high ($\log \xi \sim 2.7$)
ionization \citep[see also][for similar modeling using \texttt{Cloudy}
with \xmm/RGS data]{Armentrout2007} next to the collisionally ionized
phase. The higher ionized line-emitting gas phase detected using
\texttt{Cloudy} may also be consistent with the highly ionized warm
absorber of similar ionization ($\log \xi \sim2.8$) that we model with
\texttt{XSTAR}.

The evidence for a thermal, collisionally ionized gas phase may point
towards a contribution of a bremsstrahlung continuum to the soft
X-rays, which we, for simplicity, model with a SXCOMP continuum, that
is, scattered nuclear Comptonized emission. We are unable to favor
either of both options with our data, but provide the reader with a
short discussion on the implications of our chosen model. In this
picture, the observed that highly ionized phase, which we refer to as
a ``warm mirror''
\citep{Guainazzi2005,Guainazzi2007,Guainazzi2007proc} can act as
scattering medium for the nuclear emission. This scenario would
justify our SXCOMP component to be a long-term average with respect to
the variable nuclear HXCOMP \citep[see
also][]{Pounds1986,Yang2001,Wang2011a}. The low degree of
flux-variability that we measure between 0.6 and 1.0\,keV may be
explained with portions of this warm mirror that are located close
enough to the nuclear source to respond to its variability. In fact,
around 30\%\footnote{This number can only be approximate due to the
  recent improvements on the \chandra\ PSF.} of the soft X-rays
originate in a region that is unresolved by \chandra\
\citep{Ogle2000}. The response of this mirror at various distances
from the source could explain the lack of correlated variability
between the soft and hard bands. A correlation of the HXCOMP
variations with the prominent \ion{O}{vii} and \ion{Ne}{ix} line
fluxes can neither be claimed nor excluded with respect to the large
uncertainties at CCD resolution. Also, we can neither report to be in
favor of or against correlated variability of these lines with the
column density of the absorbers. This would be expected if the clumpy
absorber were to temporarily block the nuclear irradiation onto the
diffuse gas and promote recombination of the gas.

Other than the warm mirror, an intrinsic soft excess may also explain
the soft emission, which has phenomenologically been modeled with a
steep soft power-law \citep[e.g.,][]{Yang2001,Wang2010}. This also
involves a fit with an extra bremsstrahlung component
\citep[e.g.,][]{Warwick1995}. In a more self-consistent picture, the
soft excess emission could be provided by a combination of blurred and
unblurred ionized disk reflection, that is, by the model components
\texttt{relxillCp\_lp} and \texttt{xillver}, which we use to describe
the relativistic features inherent in the continuum. Although this
works relatively well for the soft continuum at CCD resolution,
\texttt{xillver} is unable to model the highly ionized species of the
H-like and He-like ions \ion{O}{vii}/\ion{O}{viii} and \ion{Ne}{ix},
which we observe with \xmm/RGS. In addition, \texttt{xillver} predicts
a number of strong lines from Mg, Si, and S rather than O or Ne, which
are not observed with \xmm/RGS. We therefore prefer an independent
soft continuum due to extended gas on larger scales.

\section{Conclusions and outlook}
\label{sec:conclusions}
The unique Seyfert galaxy NGC~4151 allows to both probe the
circumnuclear absorber and its strongly variable absorption together
with the effects of strong gravity close to the BH via imprints on the
reflection spectrum. As part of this work, we conducted a follow-up
study based on \citet{Keck2015} and apply the improved model
  \texttt{relxillCp\_lp} that describes blurred reflection in a
  physically motivated and self-consistent lamp-post geometry together
  with the complex set of neutral absorbers that has been frequently
  reported in the literature. We apply the resulting baseline model
to all \suzaku, \xmm,\ and \nustar\ spectra that we consider in this
work and perform a time-resolved spectral analysis of the
  neutral absorbers CA$_1$ and CA$_2$ that are variable on timescales
  from days to years. We find the soft X-rays below 1\,keV to be only
mildly variable within a maximum of $\sim$6\% as opposed to
$\sim$20\% for the HXCOMP. Strong spectral variability is
apparent between 1 and 6\,keV. As a result of our dedicated modeling, we
come up with the following conclusions:
\begin{enumerate}
\item We identify two separate point-like lamp-post components LP$_1$
  and LP$_2$ in simultaneous long-look \suzaku\ and \nustar\ spectra
  at heights of $h\sim 1.2\,r_\mathrm{g}$ and $h\sim
  15.0\,r_\mathrm{g}$, respectively. We applied the most recent model
  \texttt{relxillCp\_lp} that combines the reflection spectrum off
  each point on an ionized disk (\texttt{xillver}) with the
  appropriate relativistic transfer function. This model uses the
  Comptonization continuum \texttt{nthcomp} as primary continuum. The
  normalizations and reflection fractions of both lamp-post components
  are highly degenerate. In particular we measured a low reflection
  fraction for LP$_1$ and find that runaway pair production would
  dominate for a single and compact corona close to the BH. We
  therefore propose a vertically extended corona as opposed to two
  distinct and compact coronae. We emphasize that our results possibly
  reflect Comptonization processes in a jet-base and emphasize the
  presence of non-relativistic jets in NGC~4151. An outflowing corona
  would flatten the emissivity profile, which may explain the low
  observed reflection fraction. We must additionally consider the
  jet-base and its magnetic field as a source of synchrotron photons.
  These additional seed photons may cause the corona to be relatively
  compact in the horizontal direction. The relevance of magnetic
  fields for coronae close to the BH is strongly implied and needs to
  be carefully investigated in the future.
\item Thanks to the high count statistics provided by the long-look
  \suzaku/\nustar\ campaign, we are able to constrain a complex system
  of four separate layers of absorption, that is, two neutral absorbers,
  one of which is partially covering the nucleus with only 40\%--50\%
  (CA$_1$) and one fully covering the same (CA$_2$), a third layer of
  highly ionized absorption ($\log \xi\sim 2.8$) as constrained with
  absorption features of \ion{Fe}{xxv} and \ion{Fe}{xxvi} as well as an
  UFO with an outflow velocity of $\sim$0.16--$0.24\,c$ from
  a broad absorption feature around $\sim$8\,keV. We also showed
  that we are able to distinguish the relatively flat turnover of the
  neutral partial coverer between 3 and 6\,keV from the broad and blurred
  reflection components, which both describe a similar spectral shape
  in this energy range.
\item We observe both columns CA$_1$ and CA$_2$ to be strongly
  variable both on short timescales of 2\,d (probed with the
  long-look \suzaku\ observation) and long timescales of
  approximately one year (probed with two additional
  \suzaku\ observations as well as a \xmm\ monitoring). Both absorbers
  are responsible for the bulk of the spectral variability observed
  between 1 and 6\,keV. Their variability patterns are
  similar at all timescales, which is likely caused by the observed
  degeneracies between both columns. We are therefore unable to tell
  from our modeling, if one or even both absorbers are intrinsically
  variable. The observed column density evolution could be interpreted
  as a clumpy absorber, where one or more single clouds transit the
  line-of-sight at a distance as close as the BLR up to the inner side
  of the torus. While no BLR clouds have yet been detected in the BLR
  of NGC~4151, a clumpy nature of the inner dusty torus may still be a
  valid explanation. On the other hand, the anticorrelation of the
  irradiating photon flux with the column densities \nha\ and \nhb
  offers the alternate explanation of a radiatively driven dusty wind
  or changes in the ionization degree of the near-neutral absorber.
\item The soft X-rays below 1\,keV are only mildly variable. They
  likely originate in extended gas, which is included by the large PSF
  of \suzaku\ and \xmm. We observe a blend of unresolved emission
  lines in the CCD spectra together with a weak continuum that we
  model with the SXCOMP. We analyzed the combined \xmm/RGS spectrum of
  all seven \xmm\ pointings and the underlying soft emission
    with a phenomenological blend of Gaussian line profiles.
  Line-ratio diagnostics on the dominant He-like triplets \ion{O}{vii}
  and \ion{Ne}{ix} suggest that the gas must primarily be photoionized
  with a minor contribution of a collisionally ionized phase. Related
  to the choice of the SXCOMP as underlying continuum, we discuss the
  extended and highly ionized gas in terms of a ``warm mirror'',
  scattering nuclear Comptonized continuum emission into our line of
  sight. We outline observational evidence for this gas phase, that is,
  the $\log \xi\sim 2.8$ warm absorber and the soft line emission. The
  mild degree of variability in the soft X-rays may originate in gas
  that is located close enough to the nucleus to be able to respond to
  changes in the hard X-ray continuum within the probed timescale. We
  can exclude intrinsic soft excess emission due to blurred, ionized
  reflection.
\end{enumerate}

\begin{acknowledgements}  
  We thank the anonymous referee for providing us with a number of
  detailed comments that greatly improved the clarity of this
  manuscript. We made use of ISIS functions provided by ECAP/Remeis
  observatory and MIT
  (\url{http://www.sternwarte.uni-erlangen.de/isis/}) as well as the
  NASA/IPAC Extragalactic Database (NED), which is operated by the Jet
  Propulsion Laboratory, California Institute of Technology, under
  contract with the National Aeronautics and Space Administration. We
  thank J. E. Davis for the development of the \texttt{slxfig} module
  that has been used to prepare the figures in this work. This work
  used data obtained with the \suzaku satellite, a collaborative
  mission between the space agencies of Japan (JAXA) and the USA
  (NASA) as well as \xmm, an ESA science mission with instruments and
  contributions directly funded by ESA Member States and NASA. This
  work was supported under NASA contract No.  NNG08FD60C, and made use
  of data from the NuSTAR mission, a project led by the California
  Institute of Technology, managed by the Jet Propulsion Laboratory,
  and funded by the National Aeronautics and Space
  Administration. This research has made use of the NuSTAR Data
  Analysis Software (NuSTARDAS) jointly developed by the ASI Science
  Data Center (ASDC, Italy) and the California Institute of Technology
  (USA). A.G.M. acknowledges support from NASA grant
  NNX15AE64G. A.A.Z. has been supported in part by the Polish National
  Science Centre grants 2013/10/M/ST9/00729 and 2015/18/A/ST9/00746.
\end{acknowledgements}

 \newcommand{\noop}[1]{}

\appendix
\onecolumn

\begin{landscape} 
\section{Continuum and line parameters}
  \begin{table}[!ht]
    \caption{Summary of all \xmm-related model parameter that are left free to vary. The normalization of the Comptonized continua HXCOMP and SXCOMP are defined as unity if the norm equals one at 1\,keV. Parameters that are marked with an asterisk ($\ast$) are frozen to the parameters found for the long-look observation Suz~3.}
    \resizebox{1.3\textwidth}{!}{
      \begin{tabular}{lllllllll}
        \hline\hline 
        &  & XMM~1 & XMM~2 & XMM~3 & XMM~4 & XMM~5 & XMM~6 & XMM~7  \\
        \hline
Detconst & pn & 1$^\ast$ &1$^\ast$  & 1$^\ast$ & 1$^\ast$ & 1$^\ast$ & 1$^\ast$ & 1$^\ast$ \\
 & MOS\,1 &-- & $1.29\pm0.02$ & $1.24\pm0.02$ & $1.27\pm0.03$ & $1.27\pm 0.03$ & $1.27\pm0.06$ & $1.43^{+0.05}_{-0.04}$ \\
 & MOS\,2 &-- & $1.28\pm0.02$ & $1.28\pm0.02$ & $1.29\pm0.03$ & $1.33\pm0.04$ & $1.32\pm0.06$ & $1.32^{+0.03}_{-0.02}$ \\
HXCOMP & norm &$0.037^{+0.006}_{-0.007}$ & $0.0645\pm0.0007$ & $0.061^{+0.004}_{-0.005}$ & $0.077^{+0.005}_{-0.007}$ & $0.035^{+0.003}_{-0.006}$ & $0.055^{+0.007}_{-0.014}$ & $0.059^{+0.004}_{-0.005}$ \\
SXCOMP & norm &$\left(1.34^{+0.16}_{-0.24}\right)\times10^{-3}$ & $\left(1.42^{+0.06}_{-0.32}\right)\times10^{-3}$ & $\left(9.3\pm3.0\right)\times10^{-4}$ & $\left(1.36^{+0.10}_{-0.27}\right)\times10^{-3}$ & $\left(1.10^{+0.07}_{-0.04}\right)\times10^{-3}$ & $\left(1.3^{+0.2}_{-0.4}\right)\times10^{-3}$ & $\left(1.15^{+0.00}_{-0.05}\right)\times10^{-3}$ \\
LP~1 & norm &$\le 10$ & $\le0.4$ & $\le2$ & $\le3$ & $\le3$ & $\le6$ & $\le2$ \\
 & height [$r_\mathrm{EH}$] & 1.1$^\ast$ & \textquotedbl  & \textquotedbl  & \textquotedbl & \textquotedbl & \textquotedbl & \textquotedbl  \\
LP~2 & norm &$\left(1.6^{+1.1}_{-0.9}\right)\times10^{-3}$ & $\left(1.06\pm0.13\right)\times10^{-3}$ & $\left(1.1\pm0.5\right)\times10^{-3}$ & $\left(2.2\pm0.6\right)\times10^{-3}$ & $\left(6.7^{+3.6}_{-1.0}\right)\times10^{-4}$ & $\left(9\pm7\right)\times10^{-4}$ & $\left(1.85^{+0.42}_{-0.13}\right)\times10^{-3}$ \\
 & height~[$r_\mathrm{EH}$] & 14.1$^\ast$ & \textquotedbl  &\textquotedbl   &\textquotedbl   &\textquotedbl   & \textquotedbl  &  \textquotedbl \\
CA$_{1}$ & $N_\text{H,int}$~[$10^{22}\,\mathrm{cm}^{-2}$] &$25^{+12}_{-8}$ & $11.8^{+0.8}_{-0.7}$ & $14\pm 2$ & $17\pm4$ & $25.3^{+1.9}_{-5.1}$ & $26^{+10}_{-7}$ & $12.3^{+1.4}_{-2.4}$ \\
 & f$_\text{cvr}$ & 0.46$^\ast$ & \textquotedbl & \textquotedbl & \textquotedbl & \textquotedbl &  & \textquotedbl \\
CA$_{2}$ & $N_\text{H,int}$~[$10^{22}\,\mathrm{cm}^{-2}$] &$12.4^{+1.6}_{-0.8}$ & $6.80^{+0.11}_{-0.26}$ & $5.9\pm 0.3$ & $12.8^{+0.5}_{-0.6}$ & $15.6^{+0.4}_{-0.7}$ & $9.2^{+0.6}_{-0.7}$ & $9.05^{+0.17}_{-0.35}$ \\
Fe~K$\alpha$ & norm~[$\mathrm{Ph}\,\mathrm{s}^{-1}\,\mathrm{cm}^{-2}$] &$\left(2.4^{+0.7}_{-0.8}\right)\times10^{-4}$ & $\left(2.0^{+0.6}_{-0.4}\right)\times10^{-4}$ & $\left(2.5\pm0.6\right)\times10^{-4}$ & $\left(2.3^{+0.8}_{-0.7}\right)\times10^{-4}$ & $\left(2.8^{+0.2}_{-0.4}\right)\times10^{-4}$ & $\left(2.7^{+0.4}_{-0.9}\right)\times10^{-4}$ & $\left(1.650^{+0.621}_{-0.005}\right)\times10^{-4}$ \\
 & E~[keV] & 6.4$^\ast$ & \textquotedbl &\textquotedbl  &\textquotedbl  &\textquotedbl  & \textquotedbl & \textquotedbl \\
XSTAR\,1 & $N_\text{H}$~[$10^{22}\,\mathrm{cm}^{-2}$] &$\left(6^{+34}_{-6}\right)\times10^{21}$ & $\left(1.2\pm0.7\right)\times10^{22}$ & $\left(1.2^{+1.4}_{-1.2}\right)\times10^{22}$ & $\left(3.2^{+2.6}_{-1.8}\right)\times10^{22}$ & $\leq 8.5\times10^{21}$ & $\left(1.9^{+2.8}_{-1.8}\right)\times10^{22}$ & $\leq 9.5\times10^{21}$ \\
\hline
      \end{tabular}
    }
    \label{tab:bestfit_var_xmm}
  \end{table}

  \begin{table}[!ht]
    \caption{Summary of all \suzaku-related model parameter that are left free to vary. The normalization of the Comptonized continua HXCOMP and SXCOMP are defined as unity if the norm equals one at 1\,keV. Parameters that are marked with an asterisk ($\ast$) are frozen to the parameters found for the long-look observation Suz~3.}
    \resizebox{1.3\textwidth}{!}{
    \tiny
      \begin{tabular}{lllllllllll}
        \hline\hline 
        &  & Suz~1 & Suz~2 & Suz~2-1 & Suz~2-2 & Suz~3 & Suz~3-1 & Suz~3-2 & Suz~3-3 & Suz~3-4 \\
        \hline
        Detconst & XIS\,0 & 1$^\ast$& 1$^\ast$ &1$^\ast$  &1$^\ast$  & $0.998\pm0.005$ & $1.076^{+0.019}_{-0.018}$ & $1.010\pm0.010$ & $0.961^{+0.008}_{-0.007}$ & $0.957\pm0.008$ \\
        Detconst & XIS\,1 &$1.017\pm0.006$ & $1.032\pm0.006$ & $0.949\pm0.009$ & $1.047\pm0.006$ & $0.954\pm0.005$ & $1.043^{+0.019}_{-0.017}$ & $1.000\pm0.011$ & $0.910\pm0.007$ & $0.923\pm0.008$ \\
        & XIS\,3 &$0.995\pm0.006$ & $0.975\pm0.006$ & $0.977\pm0.009$ & $0.986^{+0.006}_{-0.005}$ & $1.012\pm0.005$ & $1.113^{+0.020}_{-0.018}$ & $1.053\pm0.011$ & $0.997\pm0.008$ & $0.993^{+0.009}_{-0.008}$ \\
        & HXD & $1.30\pm0.02$ & $1.29\pm0.02$ & $1.32\pm 0.03$ & $1.09\pm0.04$ & $1.215\pm0.009$ & $1.35\pm0.03$ & $1.239^{+0.019}_{-0.018}$ & $1.186\pm0.013$ & $1.197\pm0.014$ \\
        & FPMA &-- &--  &--  &--  & 1$^\ast$ & 1$^\ast$ & 1$^\ast$ &1$^\ast$  & 1$^\ast$ \\
        & FPMB &-- &--  &--  &--  & $1.030\pm0.004$ & $1.05\pm0.02$ & $0.998\pm0.007$ & $1.039\pm0.005$ & $1.039\pm0.007$ \\
        HXCOMP & norm &$0.0698\pm0.0013$ & $0.0726\pm0.0012$ & $0.072\pm 0.002$ & $0.055^{+0.007}_{-0.004}$ & $0.047\pm 0.002$ & $0.0406^{+0.0012}_{-0.0010}$ & $0.0481\pm0.0005$ & $0.0445\pm0.0004$ & $0.0509\pm0.0005$ \\
        SXCOMP & norm & $1.33\times 10^{-3}$\,$^\ast$ & \textquotedbl & \textquotedbl & \textquotedbl & $\left(1.33\pm 0.05\right)\times10^{-3}$ & \textquotedbl &\textquotedbl  &  \textquotedbl& \textquotedbl \\
        LP$_1$ & norm &$\left(5.3^{+1.4}_{-1.2}\right)\times10^{-4}$ & $\left(7.1^{+2.8}_{-1.9}\right)\times10^{-4}$ & $\left(6^{+4}_{-3}\right)\times 10^{-4}$ & $0.4^{+0.4}_{-0.3}$ & $6.9^{+1.2}_{-4.9}$ & \textquotedbl &  \textquotedbl& \textquotedbl & \textquotedbl \\
        & height~[$r_\mathrm{EH}$] &$24^{+9}_{-7}$ & $14^{+6}_{-4}$ & $14^{+10}_{-6}$ & $1.53^{+6.31}_{-0.15}$ & $1.1000^{+0.0013}_{-0.0000}$ & \textquotedbl & \textquotedbl & \textquotedbl & \textquotedbl \\
        LP$_2$ & norm &-- & -- & -- & -- & $\left(8\pm 2\right)\times10^{-4}$ & \textquotedbl & \textquotedbl & \textquotedbl & \textquotedbl \\
        & height~[$r_\mathrm{EH}$] &-- & -- & -- & -- & $14.1^{+3.8}_{-1.9}$ & \textquotedbl &\textquotedbl  & \textquotedbl & \textquotedbl \\
        CA$_{1}$ & $N_\text{H,int}$~[$10^{22}\,\mathrm{cm}^{-2}$] &$13^{+3}_{-2}$ & $9.2^{+1.5}_{-1.1}$ & $12^{+6}_{-4}$ & $5.6^{+1.6}_{-0.9}$ & $21^{+4}_{-3}$ & $13.7^{+4.8}_{-1.8}$ & $12.2^{+2.3}_{-1.4}$ & $24.2^{+1.9}_{-1.7}$ & $15.7^{+2.1}_{-1.4}$ \\
        & f$_\text{cvr}$ &$0.46^{+0.11}_{-0.10}$ & $0.45\pm0.06$ & $0.33^{+0.13}_{-0.08}$ & $0.49^{+0.09}_{-0.10}$ & $0.46^{+0.06}_{-0.05}$ & $0.57^{+0.13}_{-0.18}$ & $0.57\pm0.14$ & $0.51\pm0.04$ & $0.57\pm0.09$ \\
        CA$_{2}$ & $N_\text{H,int}$~[$10^{22}\,\mathrm{cm}^{-2}$] &$6.9^{+0.7}_{-0.8}$ & $4.4\pm0.3$ & $5.3^{+0.4}_{-0.6}$ & $3.9\pm 0.3$ & $8.8^{+0.5}_{-0.6}$ & $7.7^{+1.6}_{-1.3}$ & $7.3^{+1.2}_{-1.4}$ & $9.1^{+0.5}_{-0.6}$ & $7.7^{+0.9}_{-1.0}$ \\
        Fe~K$\alpha$ & norm~[$\mathrm{Ph}\,\mathrm{s}^{-1}\,\mathrm{cm}^{-2}$] &$\left(2.4\pm 0.2\right)\times10^{-4}$ & $\left(2.5\pm 0.2\right)\times10^{-4}$ & $\left(2.8\pm 0.3\right)\times10^{-4}$ & $\left(2.70^{+0.16}_{-0.14}\right)\times10^{-4}$ & $\left(2.27^{+0.13}_{-0.14}\right)\times10^{-4}$ & $\left(2.00^{+0.22}_{-0.18}\right)\times10^{-4}$ & $\left(1.75^{+0.16}_{-0.11}\right)\times10^{-4}$ & $\left(2.35^{+0.16}_{-0.09}\right)\times10^{-4}$ & $\left(2.31\pm0.14\right)\times10^{-4}$ \\
        & E~[keV] & 6.394$^\ast$ & \textquotedbl & \textquotedbl & \textquotedbl & $6.394^{+0.005}_{-0.006}$ & \textquotedbl & \textquotedbl &\textquotedbl  &  \textquotedbl\\
        XSTAR\,1 & $N_\text{H}$~[$10^{22}\,\mathrm{cm}^{-2}$] & $1.22\times10^{22}$ & \textquotedbl & \textquotedbl & \textquotedbl & $\left(1.2\pm 0.3\right)\times10^{22}$ & \textquotedbl &\textquotedbl  & \textquotedbl & \textquotedbl \\
        \hline
      \end{tabular}
    }
    \label{tab:bestfit_var_suz}
  \end{table}
\end{landscape}

\begin{table}
  \caption{Line parameters of Voigt profiles fitted to the combined RGS data of all \xmm\ observations as part of the monitoring. The parameters of the \ion{O}{vii} and \ion{Ne}{ix} lines are adopted from Table~\ref{tab:he_triplet}. The thermal velocities are unconstrained and not listed in the table. The continuum parameters are frozen to those of the SXCOMP.}
\begin{tabular}{llllll}
\hline\hline 
Line & E\,[keV] & FWHM\,[keV] & $k_\mathrm{B}\,T$ [keV] & flux\,[Ph\,s$^{-1}$\,cm$^{-2}$] \\
\hline
\ion{C}{vi}\,$\alpha$  &  $0.36637\pm0.00008$  & $\left(4.3^{+1.8}_{-1.5}\right)\times10^{-3}$  &  --  &  $\left(2.7\pm 0.3\right)\times10^{-4}$ \\
\ion{C}{v}~RRC  & $0.3901^{+0.0004}_{-0.0005}$ &  --  & $\left(4.4^{+1.4}_{-1.0}\right)\times10^{-3}$   & $\left(1.9\pm 0.3\right)\times10^{-4}$ \\
\ion{N}{vi}\,f  &  $0.41855\pm0.00009$  &  $\le1.5\times10^{-3}$  &  --  & $\left(1.35^{+0.16}_{-0.15}\right)\times10^{-4}$ \\
\ion{N}{vi}\,i  &  $0.4250\pm0.0004$  &  $\le6\times10^{-3}$  &  --  & $\left(2\pm 3\right)\times10^{-5}$ \\
\ion{N}{vi}\,r  &  $0.42909\pm0.00015$ & $\left(8^{+30}_{-8}\right)\times10^{-4}$  &  --  & $\left(9.6^{+1.6}_{-1.5}\right)\times10^{-5}$ \\
\ion{C}{vi}\,$\beta$  &  $0.4342\pm0.0002$  & $\left(7^{+50}_{-7}\right)\times10^{-4}$  &  --  & $\left(3.6^{+1.4}_{-3.6}\right)\times10^{-5}$ \\
\ion{C}{vi}\,$\gamma$  & $0.4580\pm 0.0002$ & $\le 4.8\times10^{-2}$ &  --  & $\left(3.5^{+0.5}_{-3.5}\right)\times10^{-6}$  \\
\ion{C}{vi}~RRC  & $0.4875^{+0.0005}_{-0.0006}$   &  --  & $0.010^{+0.000}_{-0.005}$  & $\left(1.5^{+0.2}_{-0.4}\right)\times10^{-4}$ \\
\ion{N}{vii}~Ly\,$\alpha$  & $0.49874^{+0.00015}_{-0.00008}$  &  $\le 2.1\times10^{-2}$ &  --  & $\left(1.6^{+0.5}_{-1.6}\right)\times10^{-3}$ \\
\ion{O}{vii}\,f & 0.55946 & $2.5\times 10^{-3}$ &  --& $4.5\times 10^{-4}$ \\
\ion{O}{vii}\,i & 0.56702 & $2.7\times 10^{-3}$ & --& $7.9\times 10^{-5}$ \\
\ion{O}{vii}\,r & 0.57200 & $2.6\times 10^{-3}$ & --& $1.3\times 10^{-4}$ \\
\ion{O}{viii}~Ly\,$\alpha$  & $0.65158^{+0.00013}_{-0.00011}$  & $\left(3.4^{+2.1}_{-1.8}\right)\times10^{-3}$ & --  & $\left(1.58^{+0.11}_{-0.12}\right)\times10^{-4}$ \\
\ion{N}{vii}~RRC  & $0.6629^{+0.0004}_{-0.0007}$  &  --  & $\left(10^{+1.2}_{-0}\right)\times10^{-3}$  & $\left(2.5^{+0.6}_{-0.5}\right)\times10^{-5}$  \\
\ion{O}{vii}~RRC  &  $0.7366\pm0.0005$  &  --  & $\left(3.0^{+0.8}_{-0.7}\right)\times10^{-3}$  & $\left(5.5\pm0.7\right)\times10^{-5}$ \\
\ion{O}{viii}~Ly\,$\beta$  & $0.7723\pm0.0006$  &  $\le4.5\times10^{-3}$  &   --  & $\left(1.6^{+0.5}_{-0.4}\right)\times10^{-5}$  \\
\ion{O}{viii}~Ly\,$\gamma$  & $0.8228\pm0.0008$  & $\le 6.3\times10^{-2}$ &-- & $0.094^{+0.007}_{-0.094}$ \\
\ion{O}{viii}~RRC  & $0.8677^{+0.0008}_{-0.0010}$   &  --  & $\left(4.0^{+1.9}_{-1.4}\right)\times10^{-3}$   & $\left(4.2^{+0.7}_{-0.6}\right)\times10^{-5}$  \\
\ion{Ne}{ix}\,f  & 0.9024   &  $\le 6\times10^{-3}$  &  --  & $5.8\times10^{-5}$ \\
\ion{Ne}{ix}\,i  & 0.9132   &  $0.041$  &  --  &  $2.2\times10^{-5}$ \\
\ion{Ne}{ix}\,r  & 0.9191   &  $\le 9\times10^{-3}$   &  --  &  $2.4\times10^{-5}$ \\
\ion{Ne}{x}~Ly\,$\alpha$  & $1.0192\pm0.0006$  & $\le9\times10^{-3}$  & --   & $\left(2.7^{+0.5}_{-1.1}\right)\times10^{-5}$ \\
\hline\hline 
Continuum & $\Gamma$ & norm [Ph\,keV$^{-1}$\,s$^{-1}$\,cm$^{-2}$] & & & \\
\hline
 & 1.72$^\ast$ & $\left(1.33\pm0.05\right)\times10^{-3\,\ast}$ & & &\\
\hline
\end{tabular}
\label{tab:lines_manualfits}
\end{table}

\end{document}